\documentclass[10pt,twocolumn,english,nofootinbib,twocolumn]{revtex4-1}
\usepackage{ae,aecompl}
\usepackage[T1]{fontenc}
\usepackage[latin9]{inputenc}
\usepackage{bm}
\usepackage{amsmath}
\usepackage{amssymb}
\usepackage{graphicx}
\usepackage{esint}
\usepackage{array}
\usepackage{booktabs }
\usepackage{topcapt}
\usepackage{enumitem}
\usepackage[unicode,breaklinks]{hyperref}
\hypersetup{
    unicode=true,
    a4paper=true,
    plainpages=false, 
    colorlinks=true,
    linkcolor=blue,
    citecolor=blue,
    filecolor=black,
    urlcolor=blue
}

\usepackage{bbold}
\usepackage{calrsfs}

\newcommand{\qt}{\tilde{q}}

\newcommand{\w}{w}
\newcommand{\xic}{\xi_c}

\newcommand{\W}{W_{\sigma}}
\newcommand{\What}{\hat{W}_{\sigma}}
\newcommand{\Sw}{S_w}

\renewcommand{\v}[1]{\ensuremath{\mathbf{#1}}} % for vectors
 
\newcolumntype{C}{>{$\displaystyle} c <{$}}

\usepackage{cases}
\usepackage{multirow}

\date{\today}
\usepackage{babel}
\begin{document}

\title{Fluctuations of spinor Bose-Einstein condensates}

\author{L. M. Symes} 
\author{D. Baillie} 
 \author{P. B. Blakie}

\affiliation{Jack Dodd Centre for Quantum Technology, Department of Physics, University
of Otago, Dunedin, New Zealand}
\begin{abstract}
We develop theory for fluctuations in atom number and spin within finite-sized cells of a spinor Bose-Einstein condensate.  This theory provides a model of measurements that can be performed in current experiments using  finite resolution  \textit{in situ} imaging.
We develop analytic results for quantum and thermodynamic limits of the fluctuations
and apply our theory to the four equilibrium phases of a spin-1 condensate.
We then validate these limits and examine the behaviour over a wide parameter regime
using numerical calculations specialised to the case of a spinor condensate confined to be quasi-two-dimensional (quasi-2D). 
\end{abstract}
\maketitle 
 
\section{Introduction}

The measurement of fluctuations can be used to  reveal important properties about equilibrium states (e.g.~see \cite{Burt1997a,Naraschewski1999,Tolra2004a,Altman2004a,Schellekens2005a,Greiner2005a,Rom2006,Trebbia2006a,Gerbier2006a,Jeltes2007a,Donner2007a,Muller2010a,Hung2011,Guarrera2011a,Hodgman2011a,Jacqmin2011a,Sanner2011a,Armijo2012a,Marzolino2013a,Blumkin2013a,Schley2013a}) and non equilibrium dynamics (e.g.~see \cite{Cheneau2012a,Hung2013}) of ultra-cold gases.
Recent experiments with trapped Bose gases have measured density fluctuations in quasi-one-dimensional (quasi-1D) \cite{Jacqmin2011a,Armijo2012a},  quasi-2D  \cite{Hung2011,Hung2011a} and three-dimensional (3D) systems \cite{Blumkin2013a,Schley2013a} using \textit{in situ} absorption imaging.   
More precisely, finite imaging resolution means that these measurements effectively count the number of atoms within a cell. The cell size strongly affects the properties of the measurement in that it selects the range of excitation wavelengths that dominant the fluctuations \cite{Klawunn2011a, Bisset2013a,Baillie2014a}. The fluctuations of multiple component systems is of increasing interest, and 
 we note theoretical work on the fluctuations of a coherently coupled two-component  condensate  \cite{Abad2013a}.

\begin{figure}[tp]
\includegraphics[width=\linewidth]{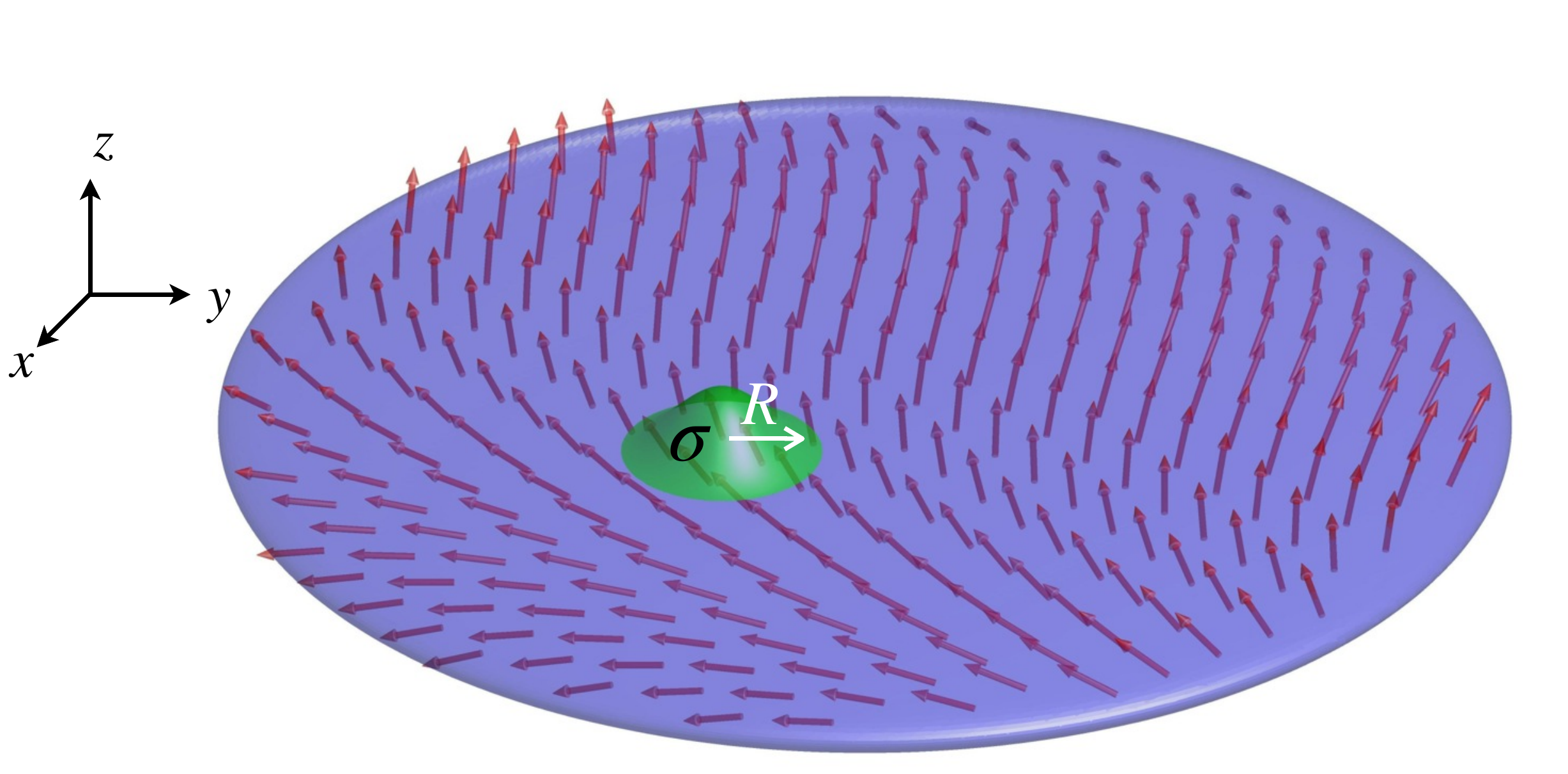}
\caption{
(Color online) Schematic illustration of fluctuation measurement for a quasi-2D spinor condensate. Within a small cell of radius $R$ [indicated by a Gaussian weight function $\sigma$ (green)] an operator of interest is measured. For example, the total number of atoms or the components of spin.
}
\label{fig:schematic}
\end{figure}

In this paper we consider the equilibrium fluctuations of spatially extended\footnote{In contrast to the tightly confined single-mode regime (e.g.~see \cite{Chang2005a}).} spinor condensates  \cite{Ho1998a,Ohmi1998a}. 
In this system the condensate  possesses a vector order parameter allowing a rich variety of accessible ground state phases \cite{Ho1998a,Ohmi1998a,Kawaguchi2012R,Stamper-Kurn2013R}, which can be explored by varying the interaction parameters and the externally applied magnetic field. Certain phases of this system exhibit multiple spontaneously broken symmetries (e.g.~see \cite{Sadler2006a}), with a matching number of Nambu-Goldstone modes. As the excitations of the system can exhibit density (e.g.~phonon excitations) and spin-density (e.g.~magnon excitations) character, it is of interest to probe the associated fluctuations in atom number and total spin within measurement cells. In practice the measurement of spin can be performed using dispersive imaging techniques \cite{Higbie2005a}  in the quasi-2D regime (see Fig.~\ref{fig:schematic}) and examples of fluctuation measurements have been reported in experiments (e.g.~see \cite{Vengalattore2010a}).
In addition to providing a general framework for calculating fluctuations we focus in on the spin-1 case, which is realised in current experiments with $^{23}$Na and $^{87}$Rb condensates. This system has four distinct magnetic phases, and for each we give analytic results for the number and spin fluctuations based upon a Bogoliubov description of the excitations. 
Our results catalog how properties of the various magnetic phases are revealed by their fluctuations.  A notable example is the broken-axisymmetric phase in which the axial spin symmetry of the Hamiltonian is spontaneously broken, causing a divergence in an axial-component of the spin fluctuations at finite temperature.  We show that in a finite system this divergence manifests as non-extensive scaling of the fluctuations.

The outline of our paper is as follows. We begin in Sec.~\ref{SEC:GenThry}  by developing a general theory for the measurement of number and spin fluctuations of a spin-$F$ condensate using a finite sized measurement cell (see Fig.~\ref{fig:schematic}). In Sec.~\ref{Sec:limits} we discuss the  thermodynamic and quantum limits of the fluctuations. 
Then in Sec.~\ref{Sec:Spin1results} we  specialise to the case of a spin-1 condensate  and present analytic results for the limiting behaviour.
In Sec.~ \ref{sec:cellFlucParams} we further specialise to the quasi-2D regime to present numerical results. We conclude in Sec.~\ref{Sec:Conclusion}.

\section{Formalism}\label{SEC:GenThry}  
Let us consider  measurements made in a cell (i.e.~localised region of space) of a spin-$F$ Bose-Einstein condensate. We take the observables to be of the form
\begin{equation}
\What\equiv\int \mathop{d^{}\mathbf{x}}\,\sigma(\mathbf{x})\hat{\w}(\mathbf{x}),
\end{equation}
where the weight function $\sigma(\mathbf{x})$ describes the  cell and $\hat{w}(\mathbf{x})$ is a generalized density operator of interest. For generality we take the spatial coordinate to be $D$-dimensional so that our formulation can be immediately specialised  to quasi-1D and quasi-2D systems.
The generalised $\hat{w}$ takes the form
\begin{equation}
\hat{w}(\mathbf{x}) \equiv \hat{\boldsymbol{\psi}}^{\dagger}\!(\mathbf{x)}\mathrm{W}\hat{\boldsymbol{\psi}}(\mathbf{x)},\label{eq:wdensityDef}
\end{equation}
where   $\mathrm{W}$ is a $(2F\!+\!1)\!\times\!(2F\!+\!1)$ matrix in spin space, and $\hat{\boldsymbol{\psi}}(\mathbf{x)}=[\hat{\psi}_F(\mathbf{x}),\hat{\psi}_{F-1}(\mathbf{x}),\ldots,\hat{\psi}_{-F}(\mathbf{x})]^T$ is a spin-$F$ bosonic field operator. For definiteness, we shall later specialise to the case of the total density  with $\mathrm{W}\to\mathbb{1}$  (the identity matrix),  and components of spin density with $\mathrm{W}\to\mathrm{F}_{x,y,z}$  (where the $\{\mathrm{F}_\alpha\}$ are the spin-$F$ spin matrices). However, we note that the formalism can also include nematic densities, although we do not present results for these here.

The variance in $\What$  is given by
\begin{equation}
\Delta \W^{2}\equiv\langle\What^2\rangle-\langle\What\rangle^{2},
\end{equation}
which can be evaluated as
\begin{align}
\Delta \W^{2} & =\!\int \!d^{}\mathbf{x}\!\int \!d^{}\mathbf{x}^\prime\,
\sigma(\mathbf{x})\sigma(\mathbf{x}')\left\langle \delta \hat{\w}(\mathbf{x}) \delta \hat{\w}(\mathbf{x}')\right\rangle,\label{d2N}%\\ 
\end{align}
where we have defined the fluctuation operator
\begin{equation}
\delta \hat{\w}(\v{x}) \equiv \hat{\w}(\v{x}) - \w.\label{e:dw}
\end{equation}
with $\w=\langle \hat{\w}\rangle$. We label the correlation function appearing in Eq.~\eqref{d2N} as
\begin{equation}
C_{w}(\mathbf{x},\mathbf{x}^\prime) \equiv \left\langle \delta \hat{\w}(\mathbf{x}) \delta \hat{\w}(\mathbf{x}')\right\rangle.
\end{equation}
The mean number of atoms in the cell is
\begin{equation}
N_\sigma =\int \mathop{d^{}\mathbf{x}}\,\sigma(\mathbf{x})\hat{n}(\mathbf{x}),
\end{equation}
where $\hat{n}(\mathbf{x})=\hat{\boldsymbol{\psi}}^{\dagger}\!(\mathbf{x)}\mathbb{1}\hat{\boldsymbol{\psi}}(\mathbf{x)}$ is the total density operator.

\subsection{Homogeneous system}
The results we derive in this paper will be for the case of a homogeneous system of total number density $n$, for which the mean cell number is 
$N_\sigma = n V_\sigma$, where
\begin{equation}
V_{\sigma}=\int \mathop{d^{}\mathbf{x}}\,\sigma(\mathbf{x}),
\end{equation}
is the  $D$-dimensional  effective volume  of the cell.
Under the assumption of homogeneity $C_{w}$ depends only on the relative separation $\mathbf{r}=\mathbf{x}-\mathbf{x}^\prime$ of the coordinates and  Eq.~\eqref{d2N} simplifies to 
\begin{align}
\Delta \W^{2} &=\!\int \!d^{}\mathbf{r} \, \tau_\sigma(\v{r}) C_{w}(\mathbf{r}),\label{e:Dw2}
\end{align}
where we have defined the \textit{geometry function}
\begin{align}
\tau_\sigma(\v{r}) \equiv \!\int \!d^{}\mathbf{x}\!\int \!d^{}\mathbf{x}^\prime\,
\sigma(\mathbf{x})\sigma(\mathbf{x}') \delta(\v{x} - \v{x}^\prime - \v{r}).
\end{align}
We can then Fourier transform (\ref{e:Dw2}) to get
\begin{align}
\Delta \W^{2} & = n\int\frac{d^{}\mathbf{k}}{(2\pi)^{D}}\Sw(\mathbf{k})\tilde{\tau}_\sigma(\mathbf{k}),\label{eq:wCellFlucsIntegral}
\end{align} 
where
\begin{equation}
\tilde{\tau}_\sigma(\mathbf{k})=\left|\int \!d^{}\mathbf{x}\, e^{-i\mathbf{k}\cdot\mathbf{x}}\sigma(\mathbf{x})\right|^2,
\end{equation}
is the Fourier space geometry function of the cell $\sigma$. 
In Eq.~(\ref{eq:wCellFlucsIntegral}) we have introduced the $\w$ static structure factor
$\Sw(\mathbf{k})$, defined as a Fourier transform of the $\w$ correlation function 
\begin{align}
S_{w}(\mathbf{k})&\equiv \frac{1}{n}\int d\mathbf{r}\,e^{-i\mathbf{k}\cdot\mathbf{r}}C_w(\mathbf{r}).\label{eq:SQdef0}
\end{align}

\subsection{Cells}
The cell geometry function $\tau_\sigma$ describes the limited resolution of measurements made in experiments. For the case of optical imaging the nature of the cell is determined by the imaging optics and the pixels used to collect the image (e.g.~see \cite{Hung2011a}). In this paper we approximate this by a Gaussian cell of the form  
\begin{equation}
\sigma(\mathbf{x})=2^{D/2}\exp(-|\mathbf{x}|^2/R^2),
\end{equation}
where the cell size $R$ is usually set by the  resolution limited spot size  \cite{Hung2011a,Jacqmin2011a}, which is typically in the range $1\,$--$\,5\,\mu$m in experiments. However, it is possible to increase this size by amalgamating the signal from multiple pixels (e.g.~see \cite{Armijo2012a}).

The Gaussian cell has an effective volume  of $V_\sigma = (2\pi)^{D/2}R^D$, and a Fourier transformed geometry function 
\begin{equation}
\tilde{\tau}_\sigma(\mathbf{k})=(2\pi R^2)^De^{-k^2R^2/2}.\label{eq:tauks}
\end{equation}
Previous theoretical treatments examining the fluctuations of ultra-cold gases  have focused on the case of \textit{hard} cells with all points in the cell (typically a $D$-dimensional sphere) equally contributing to the measurement \cite{Astrakharchik2007a,Klawunn2011a}.
While these cells have the unphysical feature of a sharp boundary, the basic behaviour of the fluctuations for the Gaussian and hard cell cases are similar for the same cell sizes.

\section{Limiting results for fluctuations}\label{Sec:limits} 

\subsection{Thermodynamic Limit (TL)}
The thermodynamic limit for fluctuations is reached for sufficiently large cells and results in fluctuations that are independent of cell details other than volume. For the case of Gaussian cells we take  \textit{large} to mean that $R$ is bigger than the microscopic length scales of the system. The relevant \textit{thermal correlation length} depends on both the phase and the measurement observable under consideration. We also require that the temperature is sufficiently high for the relevant fluctuations to be thermally activated (c.f.~the quantum limit fluctuations discussed in Sec.~\ref{Sec:Qflucts}).
In the case of scalar condensates the healing length sets both the quantum and thermal correlation lengths (e.g.~see discussion in \cite{Armijo2012a}), although different length scales can arise in the spinor case (e.g.~see Sec.~\ref{Sec:Fresults}). 

In the  large cell regime  $\tilde{\tau}_\sigma(\mathbf{k})$ becomes concentrated near $\mathbf{k}=\mathbf{0}$  [see Eq.~(\ref{eq:tauks})]. Thus the TL for fluctuations (\ref{eq:wCellFlucsIntegral}) are  \begin{align}
\Delta W_\sigma^2 & \approx  N_{\sigma}k_BT\chi_\w,\label{eq:flucThermodynamicLimit}
\end{align}
where  we have introduced
a generalised static susceptibility,
\begin{equation}
\chi_{\w}\equiv S_{\w}(0)/k_BT.
\end{equation}

\subsection{Quantum Shot Noise (QSN) }\label{Sec:QSN}
In cells much smaller than all relevant length scales, the fluctuations are dominated by the high-$k$ incoherent behaviour of the structure factor, which is given by (see Ref.~\cite{Symes2014a})
 \begin{equation}
 S_\w^\infty \equiv S_\w(k\to\infty)= \frac{1}{n}\langle \hat{\boldsymbol{\psi}}^{\dagger}\mathrm{W}^2\hat{\boldsymbol{\psi}}\rangle.\label{e:Sinfty} 
 \end{equation}
In this regime the fluctuations approach the \textit{quantum shot noise} value
\begin{equation}
\Delta \W^2 \approx N_{\sigma} S_\w^\infty. \label{eq:quantumSmallCellLimit}
\end{equation}
For the case of density fluctuations, where $\mathrm{W}\to\mathbb{1}$, this gives the familiar result of $ S_n^\infty=1$ and $\Delta N_\sigma^2 \approx N_{\sigma}$.

\subsection{Quantum Large Cell  (QLC)}\label{Sec:Qflucts}
Classically, we expect all fluctuations to vanish at zero temperature. 
However, quantum fluctuations persist. Here we consider the fluctuations at $T=0$ as measured in large cells\footnote{Small cells at temperatures $T\ll\hbar^2/MR^2k_B$ exhibit quantum shot noise fluctuations, see Sec.~\ref{Sec:QSN}.}
 where the fluctuations are dominated by the collective modes of the system.

\begin{figure}[!t]
\centering
\includegraphics[width=\linewidth]{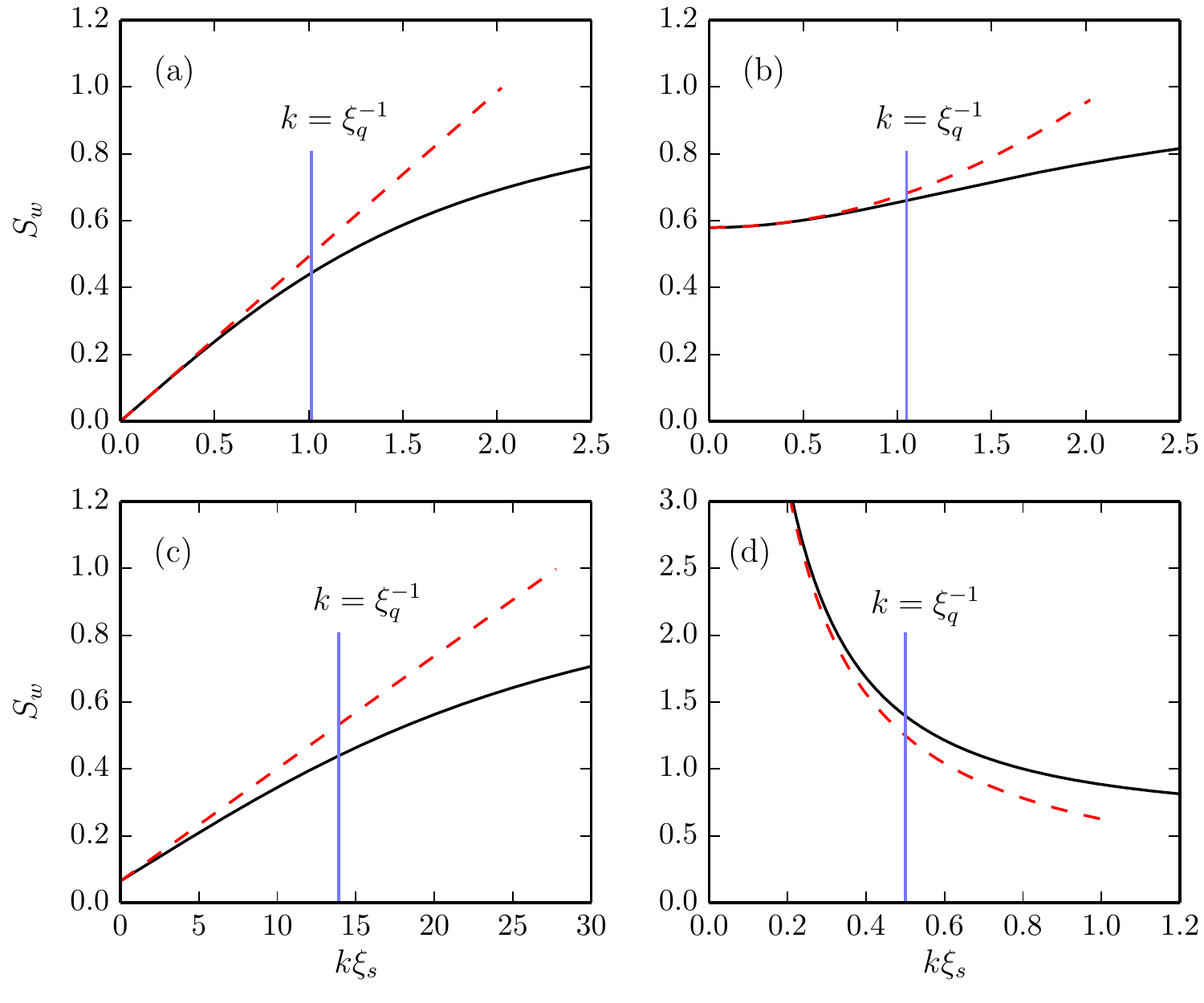}
\caption[Examples of the two models for $T=0$ structure factors.]{
\label{fig:ModelStrucs}
(Color online) Examples of $T=0$ structure factors showing the comparison to their low-$k$ expansion (\ref{EQ:SkGappedQuadratic}), along with characteristic length scale $\xic$. Subplots are:   (a) $j=1$ case of the density structure factor relevant to Sec.~\ref{Sec:AF}.  (b) $j=2$ case of the $x$-spin density structure factor relevant to Sec.~\ref{Sec:AF}. (c) $j=1$ case with $S_w(0)\ne0$ for the $x$-spin density structure factor relevant to Sec.~\ref{sec:BAcellFlucs}. For the divergent treatment presented in Eq.~(\ref{eq:Sdiverge}): (d) a $j=-1$ case  for the $y$-spin density structure factor relevant to Sec.~\ref{sec:BAcellFlucs}. 
Parameters for the results are those used in Sec.~\ref{Sec:AF} and Sec.~\ref{sec:BAcellFlucs}.
} 
\end{figure} 
 
For low-$k$ the structure factor can generally be written as
 \begin{equation}
 S_\w(k) \approx S_\w(0) + [S_\w^\infty - S_\w(0)]\left(\frac{\xic k}{2}\right)^j, \label{EQ:SkGappedQuadratic}
 \end{equation}
for $j>0$, which defines the quantum correlation length $\xic$. We show this expansion compared to examples of spinor structure factors we will use for the results developed in later sections in Fig.~\ref{fig:ModelStrucs}(a)-(c).  

Using this expansion and evaluating Eq.~(\ref{eq:wCellFlucsIntegral}) using \eqref{eq:tauks} we find the following analytic result for the quantum limit fluctuations 
\begin{align}
\frac{\Delta \W^2}{N_\sigma} &\approx S_w(0) %\notag\\&
+ [S_\w^\infty - S_\w(0)] \left(\frac{\xic}{\sqrt{2}R}\right)^j \frac{\Gamma(\frac{D+j}{2})}{\Gamma(\frac{D}{2})}. \label{eq:QLgapped}
\end{align}
where $\Gamma$ is the gamma function.
This result is valid in the limit that $R \gg \xic$.  
Interestingly, while the first term shows extensive scaling of $\Delta \W^2$ (i.e.~$\sim S_w(0)N_\sigma$), the last term exhibits non-extensive  $N_\sigma^{1 - j/D}$ scaling. 
Thus, in contrast to the TL and QSN limit, in cases where $S_w(0)=0$ the QLC fluctuations grow more slowly than $N_\sigma$.
This non-extensive nature of quantum fluctuations has been commented on in Ref.~\cite{Astrakharchik2007a}.
This result extends to small finite temperatures as long as the thermal wavelength of the relevant collective modes is larger
than the cell size \cite{Klawunn2011a} -- this is a challenging regime for experiments,
but recently there has been tremendous progress in this direction \cite{Armijo2012a}.
 
One case which does not fit Eq.~\eqref{EQ:SkGappedQuadratic} is the \textit{divergent structure factor}, where   $S_\w(k \to 0) \to \infty$, while the high-$k$ limit is non-zero, i.e. $S_\w^\infty > 0$ [e.g. see Fig.~\ref{fig:ModelStrucs}(d)]. In this case the low-$k$ behaviour can be approximated as
\begin{equation}S_\w(k ) \approx S_\w^\infty\left(\frac{ k \xic}{2} \right)^{j},  \label{eq:Sdiverge}\end{equation}
for $j<0$, which also serves to define a generalised quantum correlation length $\xic$.

Using this model, a convergent result for the fluctuations is obtained for $D > -j$, and is given by
\begin{equation}
    \frac{\Delta  {W}_{\sigma}^2}{N_\sigma} \approx S_\w^\infty \left(\frac{\xic}{\sqrt{2}R}\right)^j \frac{\Gamma(\frac{D+j}{2})}{\Gamma(\frac{D}{2})} .
\label{eq:QLdivergingSw}
\end{equation}

\section{Application to a spin-1 condensate}\label{Sec:Spin1results}
\subsection{Brief review of the spin-1 system}
In this section we specialise our theory to the case of spin-1 condensates, where the atoms can access three magnetic sub-levels $m=-1,0,1$, labelled by their $z$-projection quantum number.
The spin-1 system is conveniently described by two interaction parameters, the density-dependent coupling constant $c_0=4\pi\hbar^2(a_0+2a_2)/3M$ and the spin-depending coupling constant $c_1=4\pi\hbar^2(a_2-a_0)/3M$, where  $a_S$ ($S=0,2$) is the $s$-wave scattering length for the scattering channel of total spin $S$ and $M$ is the atomic mass \cite{Ho1998a}.
These set the corresponding density healing length $\xi_n \equiv \hbar/\sqrt{Mc_0n}$ and spin healing length $\xi_s \equiv \hbar/\sqrt{M|c_1|n}$.
The stable ground state phase is determined by the spin-dependent interaction and the linear and quadratic Zeeman energies, $p$ and $q$ respectively\footnote{Zeeman shift is $E_{\mathrm{Z}}=-pm+qm^2$.  We emphasise that $q$ can be adjusted independently of $p$, e.g.~\cite{Gerbier2006}, and that $p$ also acts as a Lagrange multiplier to constrain the $z$-component of magnetisation. } associated with an effective magnetic field along $z$.

\begin{figure}[tbhp]
   \centering
   {\renewcommand{\arraystretch}{1.7}
\begin{tabular}{c | p{4.8cm} |c}
\hline Phase & Order parameter $\boldsymbol{\xi}$ and magnetization & $\Psi(\hat{s})$
\\
\hline F & $\left[1, 0, 0 \right]^T$ or $\left[0, 0, 1 \right]^T$. \vspace*{0.5cm} \newline Fully magnetized $|f_z|=n$, $f_\perp=0$. \newline & \raisebox{-0.6\height}{\includegraphics[width=2.1cm, trim=0 0 0 -3]{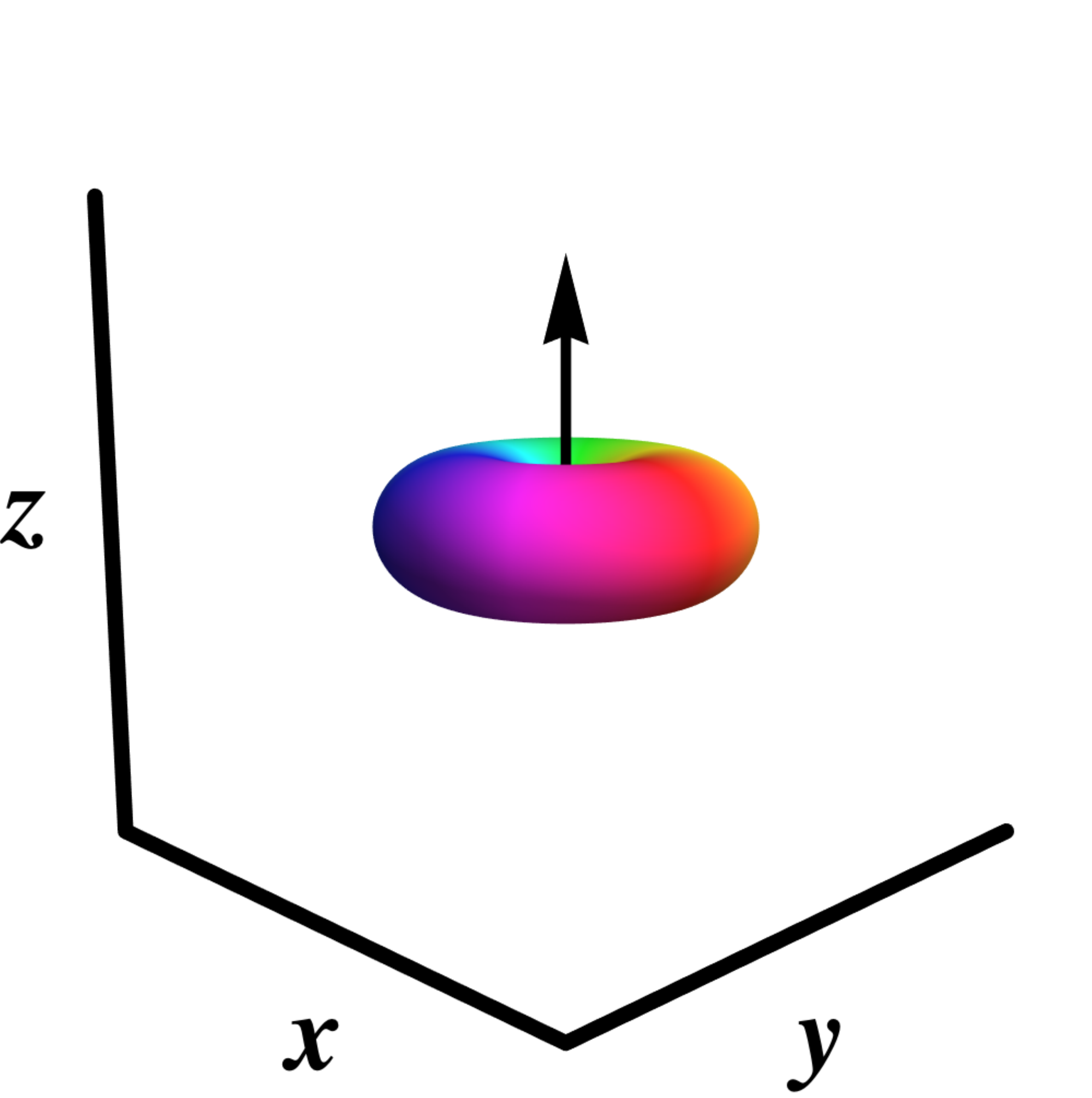}}
\\
\hline P & $\left[0, 1, 0 \right]^T$. \vspace*{0.5cm} \newline Unmagnetized $f_z=0$, $f_\perp=0$. \newline & \raisebox{-0.6\height}{\includegraphics[width=2.1cm, trim=0 0 0 -5]{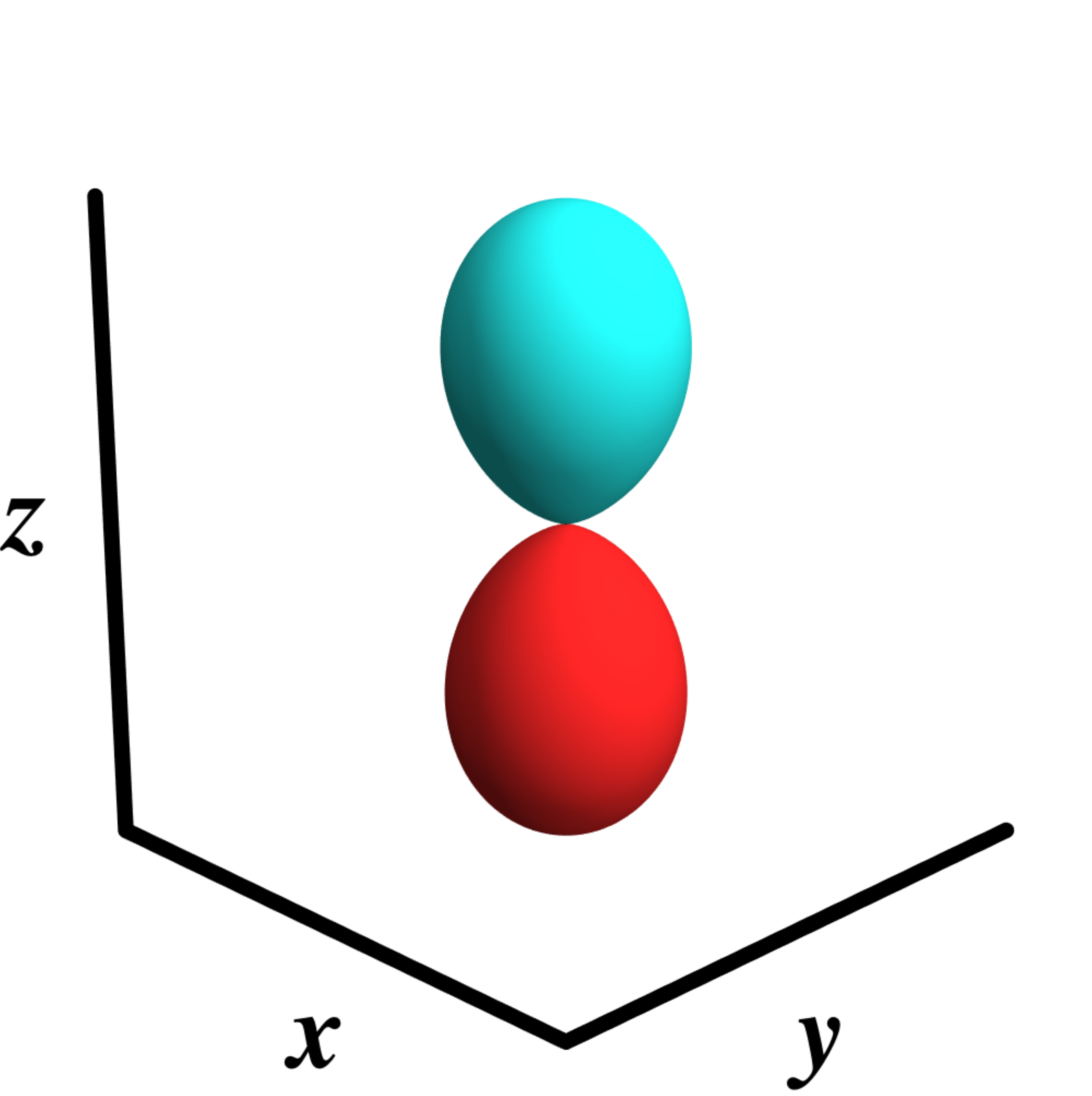}}
\\
\hline AF & $\left[\sqrt{\tfrac{1}{2}\left(1+\frac{{f}_z}{n}\right)}, 0, \sqrt{\tfrac{1}{2}\left(1-\frac{f_z}{n}\right)} \right]^T$. \newline\newline Partially magnetized $|f_z|<n$, $f_\perp=0$. & \raisebox{-0.6\height}{\includegraphics[width=2.1cm, trim=0 0 0 -5]{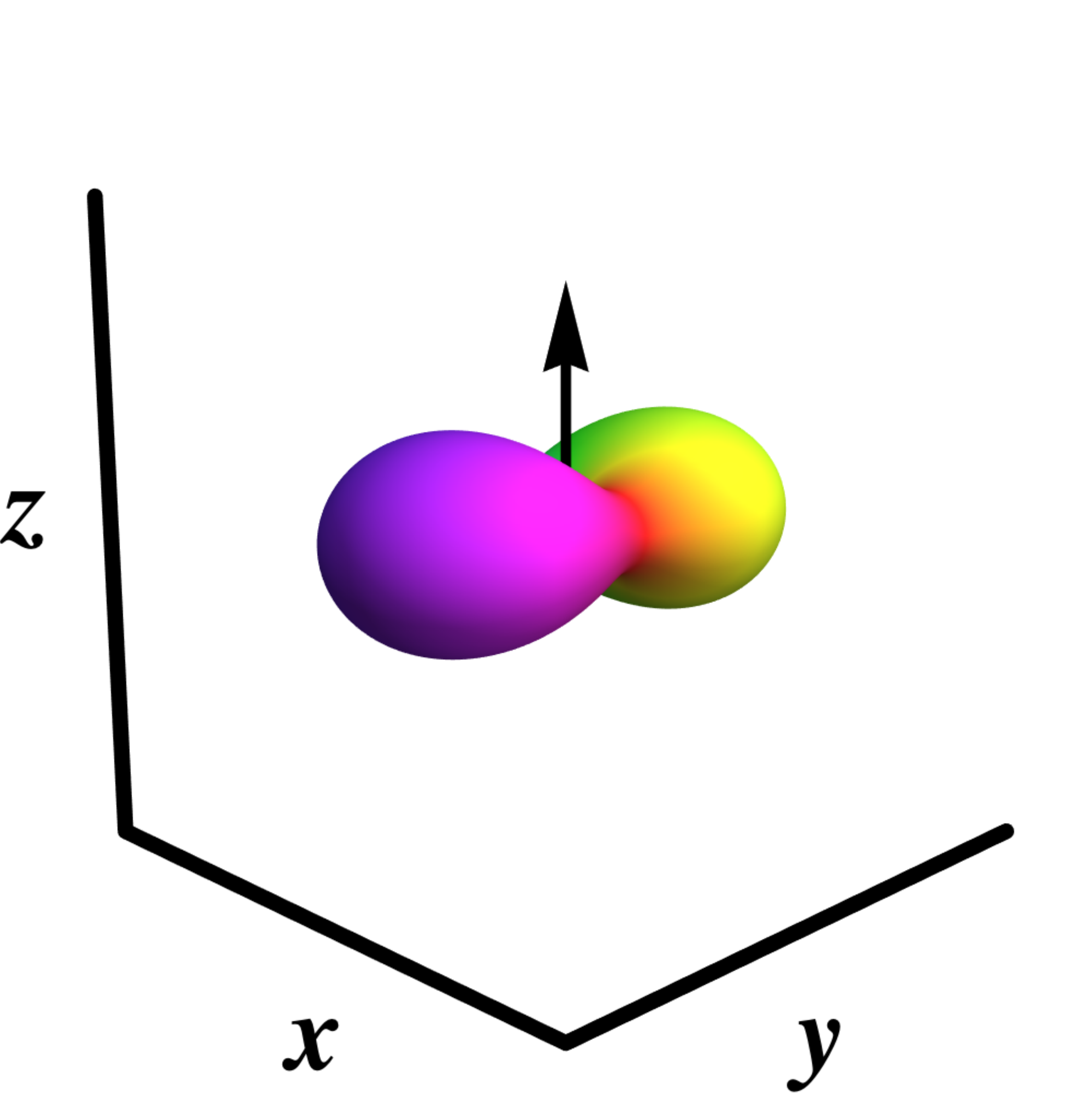}}
\\
\hline \multirow{1}{1.2cm}[-0.4cm]{\centering BA $(p=0)$} & $\left[\tfrac{1}{2}\sqrt{1 - \tilde{q}}, \sqrt{\tfrac{1}{2}(1 + \tilde{q})}, \tfrac{1}{2}\sqrt{1 - \tilde{q}} \right]^T$. \newline \newline Magnetized in-plane $|f_z|=0$, $f_\perp>0$. & \raisebox{-0.6\height}{\includegraphics[width=2.1cm, trim=0 0 0 -5]{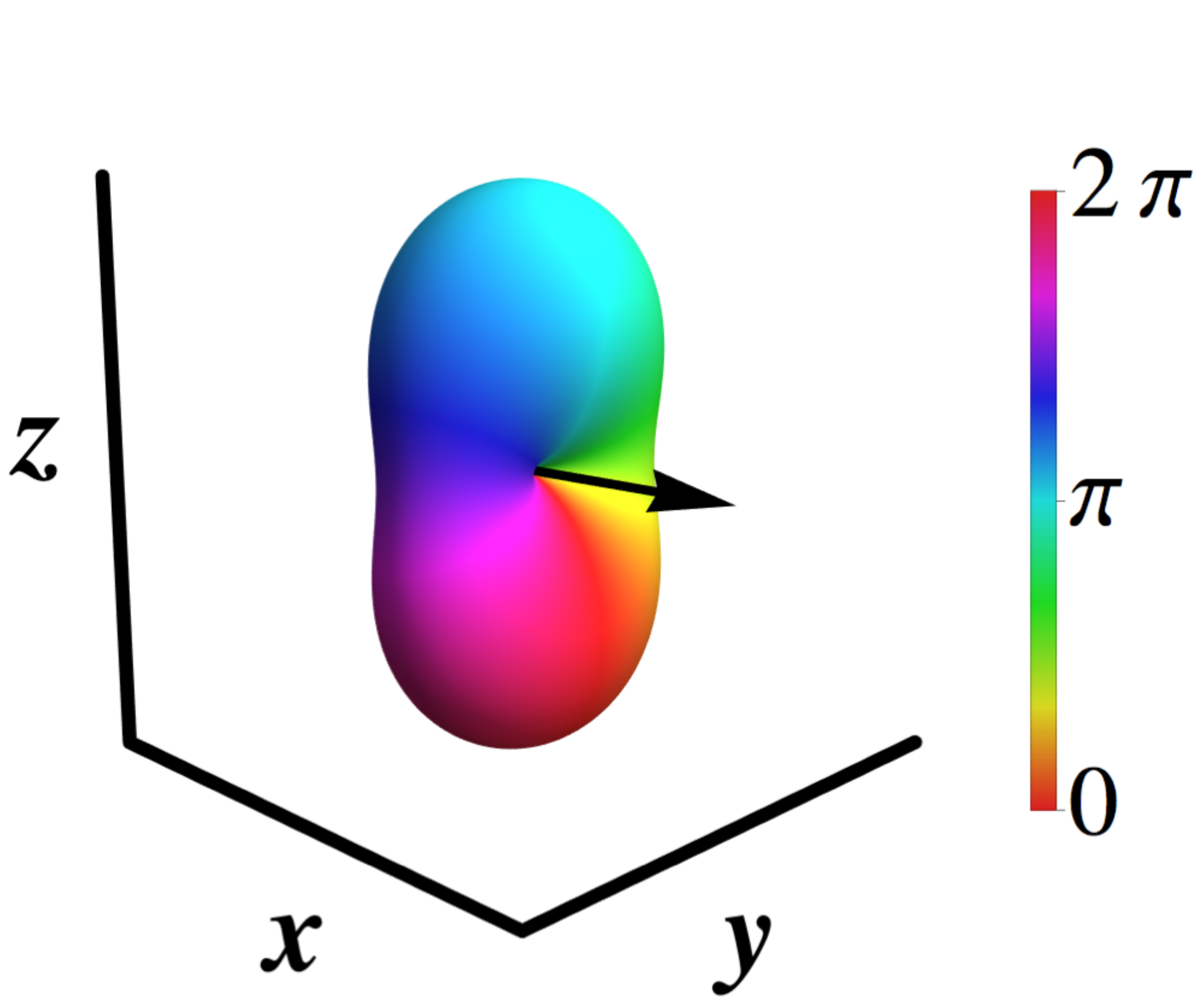}}
\\
\hline 
\end{tabular}
}
\caption{(Color online) Phases of a uniform spin-1 condensate.   The normalised spinor, $\boldsymbol{\xi}=[\xi_{1},\xi_{0},\xi_{\!-\!1}]^{T}$, magnetisation characteristics and spherical-harmonic representation of the four distinct magnetic phases.  The spherical-harmonic representation figures are surface plots of $|\Psi(\hat{s})|^2$, with the surface color  indicating $\mathrm{arg}[\Psi(\hat{s})]$, and the arrow indicates the direction of magnetisation (if any). For the BA phase we have restricted our attention to the  $p=0$ case where the magnetisation is in-plane.  \label{Tab:phases}}
\end{figure}

The ground state phase is characterized by the normalised spinor  $\boldsymbol{\xi}$ obtained from the condensate order parameter 
\begin{equation}
\langle\hat{\boldsymbol{\psi}}\rangle=\sqrt{n}\,\boldsymbol{\xi}.
\end{equation} 
A full review of the ground state phases and their properties is too lengthy to include here, and is covered comprehensively in the recent review article of Kawaguchi \textit{et al.}~\cite{Kawaguchi2012R}. However, we briefly introduce the phases in Fig.~\ref{Tab:phases}.
The four phases are the (F) ferromagnetic, (P) polar, (AF) antiferromagnetic,
 and broken-axisymmetric (BA) phases. These are distinguished by their magnetization, both along the direction of the external field (i.e.~$f_z$) and perpendicular to it (i.e.~$f_\perp \equiv\sqrt{f_x^2+f_y^2}$), where $f_\alpha= n\boldsymbol{\xi}^{\dagger}\mathrm{F}_{\alpha}\boldsymbol{\xi},$ 
is the $\alpha$-component of the condensate spin density and $\{\mathrm{F}_{\alpha =x,y,z}\}$ are the spin-1 matrices. 
Using a spherical-harmonic representation, the spinor order parameters are conveniently visualised as a complex wave function $\Psi(\hat{s})=\sum_m\xi_mY^m_1(\hat{s})$, where $\{Y^m_1(\hat{s})\}$ are the degree 1 spherical harmonics and $\hat{s}$ is a unit vector in spin-space \cite{Kawaguchi2012R}. These visualisations of the order parameter are also presented in Fig.~\ref{Tab:phases}. While the Hamiltonian for the system is axially symmetric in spin-space about $z$ (i.e.~the magnetic field direction), the order parameters of the AF and BA phases break axial symmetry.  This broken symmetry is reflected in these phases  developing new Nambu-Goldstone modes, i.e.~gapless magnon excitation branches (e.g.~see \cite{Kawaguchi2012R}).

\subsection{General results for the spin-1 system}
In this section we introduce the fluctuation observables of interest, briefly review our formalism for calculating the structure factors, and then present analytic results that can be used in conjunction with the limiting results of Sec.~\ref{Sec:limits}. 
 
The observables of interest here are the  total density and the three components of spin density. We specialise the general notation introduced earlier to these cases according to 
\begin{subequations}
\begin{align}
\hat{\w} &\to \{\hat{n},\hat{f}_x,\hat{f}_y,\hat{f}_z\}, \label{e:winterest}\\
\What &\to \{\hat{N}_\sigma, \hat{F}_{x,\sigma}, \hat{F}_{y,\sigma}, \hat{F}_{z,\sigma}\}, \label{eq:wHatFlucs}\\
S_w(\mathbf{k}) &\to \{S_n(\mathbf{k}),S_x(\mathbf{k}),S_y(\mathbf{k}),S_z(\mathbf{k})\}, \\
\chi_w &\to \{\chi_n, \chi_x, \chi_y, \chi_z\},
\end{align}
\end{subequations}
where the total density operator $\hat{n}$ was introduced earlier, and 
$\hat{f}_{\alpha}(\mathbf{x})   =   \hat{\boldsymbol{\psi}}^{\dagger}\!(\mathbf{x})\mathrm{F}_{\alpha}\hat{\boldsymbol{\psi}}(\mathbf{x})$ are the spin-density operators.

 \begin{table*}[tbp!]
   \centering
   {\renewcommand{\arraystretch}{1.7}
   \footnotesize
   \begin{tabular}{
   @{} >{\centering\arraybackslash} p{1.05cm} |
   >{\centering\arraybackslash} p{1.4cm} |
   >{\centering\arraybackslash}p{0.7cm}
   >{\centering\arraybackslash}p{5.6cm}
   >{\centering\arraybackslash}p{2.9cm}
   >{\centering\arraybackslash}p{1.5cm} |
   @{} >{\centering\arraybackslash} p{3.9cm}} % Column formatting, @{} suppresses leading/trailing space
      \hline
      \bf{Phase} & \bf{Obs.(s)} &  \multicolumn{4}{c|}{\bf{Quantum Limits}}  & \bf{Thermodynamic Limit} \\
       & $\hat{W}_\sigma$ & $j$ & $\xic$ & $S_w(0)$ & $S_w^\infty$
        & $\chi_w$ \\
		\hline
        {\centering F} & $\hat{N}_\sigma, \hat{F}_{z,\sigma}$ & $1$  & $\dfrac{\xi_n}{\sqrt{1+c_1/c_0}}$ & $0$ & $1$& $1/(c_0 + c_1)n$ \\
		 		 & $\hat{F}_{x,\sigma}, \hat{F}_{y,\sigma}$ &    &  & $1/2$ & $1/2$& $1/E_{g,2}^\mathrm{F}$ \\
		          \hline
		{\centering P} & $\hat{N}_\sigma$ & $1$ & $\xi_n$  & $0$ & $1$& $1/c_0 n$ \\
		& $\hat{F}_{z,\sigma}$ &   &  & $0$ & $0$ & $0$ \\
        & $\hat{F}_{x,\sigma}, \hat{F}_{y,\sigma}$ & $2$ & $l_\mathrm{P}^{\mathrm{ave}}\sqrt{1+\sqrt{\dfrac{q}{q + 2 c_1n}}}$  & $\sqrt{\dfrac{q}{q + 2c_1n}}$ & $1$& $\dfrac{2q}{q(q+2c_1n)-p^2}$ \\
		\hline
		& & & & & & \\[-0.4cm]
      {\centering AF} & $\hat{N}_\sigma$ & $1$ & $\xi_n\left[1-\dfrac{f_z^2}{n^2}\left(\dfrac12\dfrac{c_1}{c_0}  - \sqrt{1-\dfrac{f_z^2}{n^2}} \dfrac{c_1^{3/2}}{c_0^{3/2}}\right)\right] $ & $0$ &  $1$& $1/c_0 n$\\[0.5cm]
     % & $C=7.4$ & &\\
      & $\hat{F}_{z,\sigma}$ & $1$ & $\xi_s\!\left[\sqrt{1-\dfrac{f_z^2}{n^2}}\left(1-\dfrac{3f_z^2}{2n^2}\dfrac{c_1}{c_0} \right) +  \dfrac{f_z^2}{n^2}\sqrt{\dfrac{c_1}{c_0}} \right] $ & $0$ &  $1$& $1/c_1 n$\\[0.5cm]
      & $\genfrac{}{}{0pt}{}{ \hat{F}_{x,\sigma} \to + }{\hat{F}_{y,\sigma}\to - }$
      & $2$ & $l_g^\mathrm{AF} \sqrt{\dfrac{\vphantom{E}2\alpha_z }{\alpha_z \pm (1 - q/c_1n - E_{g,2}^\mathrm{AF}/c_1n)}}$
      & $\frac{1}{2}\left[\dfrac{(1\pm\alpha_z)E_{\mathrm{g},2}^{\mathrm{AF}}}{(1\pm\alpha_z)c_1 n - q} \right]$
      &  $\frac{1}{2}(1\pm\alpha_z)$& $\dfrac{1\pm\alpha_z}{(1\pm\alpha_z)c_1n-q}$ \\[0.5cm]
         \hline 
	  {\centering BA $(p=0)$} & $\hat{N}_\sigma$ & $1$ & $\dfrac{\xi_n}{\sqrt{1+c_1/c_0}}$ & $0$ &  $1$& $1/(c_0 + c_1)n$\\
      & $\hat{F}_{x,\sigma}$ & $1$ & $\dfrac{\xi_n}{[1-\qt^2(1+\sqrt{1-\qt^2})]\sqrt{1+c_1/c_0}}$ & $\dfrac{\qt^2}{\sqrt{1-\qt^2}}$ &  $1$& $\dfrac{1}{1 - \tilde{q}^2}\!\! \left[ \dfrac{1}{(c_0 + c_1)n} + \dfrac{	\tilde{q}^2}{|c_1|n}\right]$\\
      & $F_{y,\sigma}$ & $-1$ & $\sqrt2 \xi_{q}$ & $\infty$ &  $\frac{1}{2}(1 + \tilde{q})$& $\infty^\star$ \\
      & $\hat{F}_{z,\sigma}$ & $1$ & $\sqrt2 \xi_{q}$ & $0$ &  $\frac{1}{2}(1 - \tilde{q})$& $(1 - \tilde{q})/q$\\
	    \hline
	    \hline
       \end{tabular}
   }
   \caption{
   Analytic results for cell measurement fluctuations in a uniform spin-1 condensate.
   We detail parameters for all four magnetic phases needed to evaluate the QLC limit
   [Eq.~\eqref{eq:QLgapped} for $j>0$, Eq.~\eqref{eq:QLdivergingSw} for $j<0$], QSN limit [Eq.~\eqref{eq:quantumSmallCellLimit}],
   and the TL [Eq.~\eqref{eq:flucThermodynamicLimit}] for the observables in Eq.~\eqref{eq:wHatFlucs}.
   Parameters for AF and BA are $\tilde{q} \equiv q/2|c_1|n$, $\alpha_z \equiv \sqrt{1-(f_z/n)^2}$ and we define the quadratic Zeeman length scale $\xi_{q} \equiv \hbar/\sqrt{M q}$. 
   The quantities $E_{g,2}^\mathrm{F}\equiv p-q$ and $E_{g,2}^\mathrm{AF} \equiv c_1n\sqrt{(1-q/c_1n)^2-\alpha_z^2}$ are energetic gaps of particular excitation branches, with the notation chosen to match Ref.~\cite{Symes2014a}. 
  We have defined the length scale
   $l_\mathrm{P}^{\mathrm{ave}} \equiv \hbar/\sqrt{M (E^\mathrm{P}_{g,1} + E^\mathrm{P}_{g,2})/2} 
   = \hbar/\sqrt{M \sqrt{q(q+2c_1n)}}$  using the mean of the  energy gaps for the two P phase magnon branches (given in Sec.~\ref{Sec:AF}), and
   $l_\mathrm{AF} \equiv \hbar/\sqrt{M E_{g,2}^\mathrm{AF}}$ using the energy gap of the AF phase magnon branch.
   For $\hat{N}_\sigma$ and $\hat{F}_{z,\sigma}$ in the AF phase the expressions for $\xic$ are the first terms in an expansion for $c_1\ll c_0$.
   \\
   $^\star$ we cover the special case of the diverging $\hat{F}_{y,\sigma}$ TL separately in Sec.~\ref{sec:BAcellFlucs}.
   \label{tab:cellFlucLimits}
   }
\end{table*}

For a highly condensed system, the field operator takes the form 
$\hat{\boldsymbol{\psi}}(\mathbf{x})=\sqrt{n}\boldsymbol{\xi}+\hat{\boldsymbol{\delta}}(\mathbf{x})$, where $\hat{\boldsymbol{\delta}}(\mathbf{x})$ describes the fluctuations on the condensate.
A Bogoliubov transformation to a quasiparticle basis can be applied to $\hat{\boldsymbol{\delta}}(\mathbf{x})$ to diagonalise the many-body Hamiltonian expanded to quadratic order in $\{\hat{\boldsymbol{\delta}},\hat{\boldsymbol{\delta}}^\dagger\}$. Because of the spin degrees of freedom there are three branches of quasiparticles for the spin-1 system, which can be characterised as phonon and magnon branches according to their effect on the condensate density or magnetisation (e.g.~see \cite{Murata2007a,Uchino2010a,Kawaguchi2012R,Yukawa2012a,Symes2014a}). In terms of these excitations the $w$-density fluctuation operator (\ref{e:dw}) is given to leading order as

\begin{equation}
    \delta \hat{w}(\mathbf{x})\approx\sqrt{n}\left(\boldsymbol{\xi}^\dagger\mathrm{W}\hat{\boldsymbol{\delta}}+\hat{\boldsymbol{\delta}}^{\dagger}\mathrm{W} {\boldsymbol{\xi}}\right).\label{e:dwBog}
\end{equation} 
The $w$-structure factor is then given  using Eq.~(\ref{eq:SQdef0}).

Evaluating the $\w$-structure factor hence requires full knowledge of the condensate order parameters, quasiparticle amplitudes and excitation  energies. In general this must be done numerically, however in recent work we have obtained analytic expressions for the limiting behaviour of the structure factors  for all four phases of the spin-1 system and for the observables of interest [i.e.~Eq.~(\ref{e:winterest})] (see Table II of \cite{Symes2014a}). Utilizing this information we can evaluate  the limiting results discussed in Sec.~\ref{Sec:limits}, in particular the generalised susceptibility $\chi_w$ and the quantum correlation length $\xic$, for most phases.  These results are summarised in Table~\ref{tab:cellFlucLimits} and form a key result of this work.

\section{Calculations for the Quasi-2D regime} \label{sec:cellFlucParams}

To illustrate the applicability of our analytic results for the limiting behavior, we have evaluated the fluctuations of a spin-1 condensate using numerical calculations. 
We choose to do this for the quasi-2D regime\footnote{Condensation is  forbidden in an infinite 2D system \cite{Wagner1966,Hohenberg1967}. However, our results will apply to finite systems at sufficiently low temperatures (see Sec.~\ref{sec:BAcellFlucs}).  }  (i.e.~$D=2$), which can be realised with a  harmonic trap in which one direction, which we take to be $z$, is much tighter than the other directions. High numerical aperture \textit{in situ} imaging can then be used to realise highly resolved measurement cells in the condensate (e.g.~see Fig.~\ref{fig:schematic}). 
Indeed, detailed fluctuation measurements of this type have been performed on a scalar condensate \cite{Hung2011}, and this is approximately the regime of the experiments carried out on $^{87}$Rb spinor gases by the Berkeley group (e.g.~see \cite{Higbie2005a,Sadler2006a,Vengalattore2008a,Vengalattore2010a,Guzman2011a}). Within the quasi-2D approximation the tightly confined direction can be eliminated by integrating it out, assuming that the condensate profile in that direction is well approximated by a harmonic oscillator ground state. In doing this quasi-2D interaction parameters are obtained $c_{0,1}^{2D}=c_{0,1}/\sqrt{2\pi}l_z$, where $l_z=\sqrt{\hbar/M\omega_z}$ is the $z$ oscillator length and $\omega_z$ is the respective trap frequency.  The results of Table~\ref{tab:cellFlucLimits} cover the homogeneous (i.e.~neglecting any $x$ and $y$ confinement) quasi-2D regime with the replacements $c_{0,1}\to c_{0,1}^{2D}$, and taking $n$ to refer to the areal density of the condensate. For this reason in the next subsections references to earlier results involving interaction terms (i.e., $c_{0,1}n$) will be taken to mean the quasi-2D equivalents. The numerical results we develop are also evaluated within the homogeneous assumption. This should provide a good description of the central region of a harmonically trapped system where the mean density is almost constant.
 \begin{table}[!tbp]
\centering
\renewcommand{\arraystretch}{1.4}
\begin{tabular}{>{\centering\arraybackslash} p{2cm} | >{\centering\arraybackslash} p{3cm} | >{\centering\arraybackslash} p{3cm}}
\hline
Parameter & $^{23}$Na & $^{87}$Rb \\
\hline
$n$ & $6.1\times 10^9$ cm$^{-2}$ & $1.2\times 10^{10}$ cm$^{-2}$ \\
$c_0^{2D} n/k_B$ & $14$ nK & $28$ nK \\
$|c_1^{2D}| n/k_B$ & $0.28$ nK & $0.11$ nK \\
\hline
$\xi_n$ & $1.2$ $\mu$m & $0.45$ $\mu$m \\
$\xi_s$ & $8.7$ $\mu$m & $7.1$ $\mu$m \\
\hline\hline
\end{tabular}
\caption[Species-specific parameters for numerically calculating fluctuations within cells.]{
Species-specific parameters for numerically calculating fluctuations within cells.
Note that for $^{87}$Rb the spin dependent interaction is negative. 
\label{tab:cellFlucParams}
}
\end{table}

For definiteness, we choose physical parameters for our calculations that could be realised in current experiments:  the central part of a condensate of $N=10^6$ atoms in a 3D harmonic trap with $\omega_z=2\pi\times300$ s$^{-1}$ and radial confinement of $\omega_{x,y}=2\pi\times5$ s$^{-1}$. The ideal condensation temperature for this system is $T_c \approx100$ nK. 
For the F, P and AF phases, we use $^{23}$Na interaction parameters (i.e. $c_1>0$ noting that the F and P phases also occur for $c_1<0$),
while for the BA phase (which requires $c_1<0$) we use  $^{87}$Rb parameters, as listed in Table~\ref{tab:cellFlucParams}.

In the following subsections, we compare our numeric and analytic results for the fluctuations of the measurement operators given in
Eq.~\eqref{eq:wHatFlucs} in each ground state phase.
We discuss the fluctuation scaling with temperature and cell size. For our results we consider temperature ranges that exceed the value of $T_c$ for example physical system discussed above.  We do this to illustrate the asymptotic scaling of the fluctuations, but emphasise that our results are only applicable to particular systems below the critical temperature (also see discussion in Sec.~\ref{sec:BAcellFlucs}).

\subsection{F phase}\label{Sec:Fresults}
\begin{figure}[t]
\includegraphics[width=\linewidth]{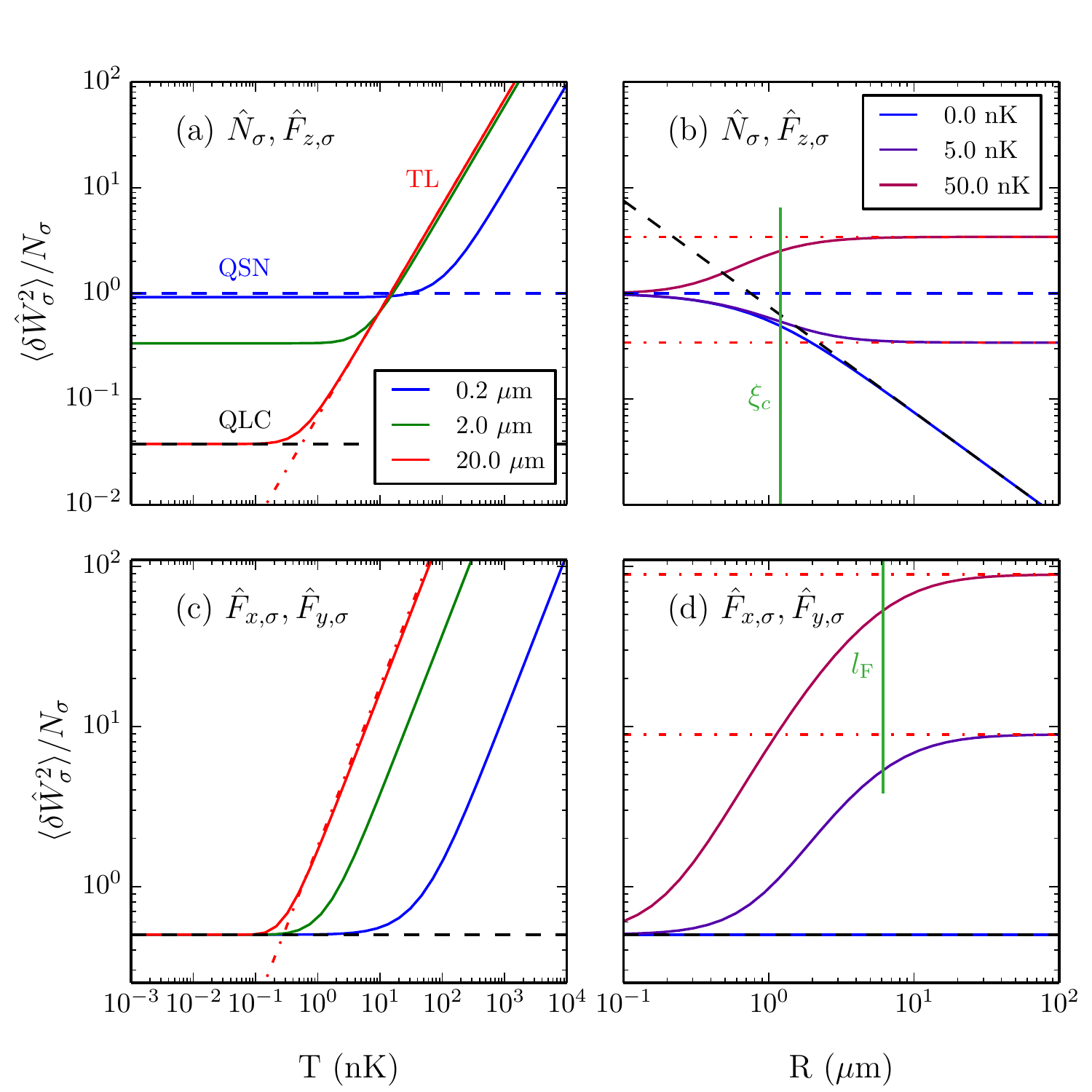}
\caption{
(Color online) Cell fluctuations for the F phase.
In (a) and (c) we plot temperature dependent results for
cell sizes of $R\in\{0.2, 2, 20\} \times \mu$m, as labelled in (a).
In (b) and (d) we plot fluctuations as a function of cell size $R$ for temperatures $T \in \{0, 5, 50\}$ nK, as labelled in (b). 
Also plotted are the QSN limit (dashed blue line), QLC limit (dashed black line) and the TL (dot-dashed red line).
Parameters are for $^{23}$Na as in Table~\ref{tab:cellFlucParams}, and we also set $f_z=n$, $p=c_1 n$ and $q=-c_1n$.
For $\hat{N}_\sigma$ and $\hat{F}_{z,\sigma}$, $\xic = 1.2$~$\mu$m [vertical green line in (b)].
For $\hat{F}_{x,\sigma}$ and $\hat{F}_{y,\sigma}$, $l_{\mathrm{F}}$ is shown [vertical green line in (d)] and $\xic = 0$.
} 
\label{fig:F-flucs}
\end{figure}

We focus on the F phase with $f_z=n$ and note that results for the case $f_z=-n$ (i.e.~$\boldsymbol{\xi}=[0,0,1]^T$) are equivalent. Numerical results are presented, along with analytic limits, in Fig.~\ref{fig:F-flucs} as functions of temperature $T$
and cell size $R$.

We first consider fluctuations in $\hat{N}_\sigma$ and $\hat{F}_{z,\sigma}$, which are presented in Figs.~\ref{fig:F-flucs}(a) and (b).
We note that, to our level of approximation for the structure functions [i.e.~Eq.~(\ref{e:dwBog})], the fluctuations in these operators are identical in this phase.
This is because the leading order effect arises from the phonon excitations in the $m=1$ sub-level (where the condensate entirely resides), and these excitations have an identical effect on both operators.
Furthermore, the structure factor has $S_\w(0)=0$ and $S_\w^\infty=1$, with linear low-$k$ behaviour,
and so the fluctuations in $\hat{N}_\sigma$ and $\hat{F}_{z,\sigma}$ are equivalent to the  number fluctuations of a scalar condensate,
which has been studied thoroughly \cite{Klawunn2011a} (also see \cite{Armijo2012a}).
Thus the analysis we now give for these operators effectively reiterates those findings.

The QSN limit is Poissonian
\begin{equation}
\Delta F_{z,\sigma}^2=\Delta N_\sigma^2=N_\sigma,
\end{equation}
[from Eq.~(\ref{eq:quantumSmallCellLimit}) and Table~\ref{tab:cellFlucLimits}] and occurs  for $R\ll \xic$, and at temperatures sufficiently low that excitations with wavelength comparable to the cell size are not thermally activated. In Fig.~\ref{fig:F-flucs}(a) this is observed for the $0.2\mu$m cell at  $T\ll500$ nK,  where the thermal wavelength  $l_T=\hbar/\sqrt{Mk_BT}$  is comparable to the cell size (see discussion in \cite{Klawunn2011a}). In  Fig.~\ref{fig:F-flucs}(b) the results for all three values of $T$ are seen to approach the QSN limit for values of $R$ much less than $\xic$.

Because the structure factor for $\hat{N}_\sigma$ and $\hat{F}_{z,\sigma}$ is ungapped with linear low-$k$ behaviour, the  QLC  limit for the quasi-2D system is [from Eq.~\eqref{eq:QLgapped}] 
\begin{equation}
\Delta F_{z,\sigma}^2=\Delta N_\sigma^2=N_\sigma\sqrt{\frac{\pi}{8}} \frac{\xic}{R}.
\end{equation}
In Fig.~\ref{fig:F-flucs}(a) this limit is only obtained for the largest cell considered (the only cell with $R\gg\xic$) at temperatures below $T\approx0.1$ nK, where there is insufficient thermal energy to activate phonon modes of wavelengths comparable to the cell size.\footnote{In \cite{Klawunn2011a} this is equivalently cast as the thermal phonon wavelength being larger than the cell size (also see \cite{Armijo2012a}).} In Fig.~\ref{fig:F-flucs}(b) the non-extensiveness of the quantum  fluctuations is revealed by the decrease in  $T=0$ fluctuations (relative to $N_\sigma$) as $R$ increases.

The TL behaviour is revealed in Fig.~\ref{fig:F-flucs}(a) by the linear growth in fluctuations with $T$ at sufficiently high $T$. The fluctuations for the smallest cell lie appreciably below the results for the large cells. This is because $\xic\gg R$ for the smallest cell, thus correlation effects are important even at high temperatures (also see discussion in Sec.~IV.B of Ref.~\cite{Klawunn2011a}).

Due to the axial symmetry of the F phase, we have that $\Delta F^2_{x,\sigma} = \Delta F^2_{y,\sigma}$. For these fluctuations, the associated structure factor has the correlation length $\xic=0$. This is because the relevant quasiparticle modes are magnons with a free particle dispersion relation, but with an energy gap $E_{g,2}^\mathrm{F} \equiv p-q$ (i.e.~set by the Zeeman energies).\footnote{In experiments $p$ is typically large ($p\sim h\times10$ kHz) and these excitations will be frozen out at condensation temperatures. We have chosen to use a small value of $p$  to illustrate transverse fluctuations at low $T$.} This means that the QLC and QSN limits are the same, since at $T=0$ the structure factors $S_{x,y}(k)=\frac{1}{2}$ are independent of $k$ (see Sec.~IV.A.3 of \cite{Symes2014a}). The equivalence of these limits is seen in Fig.~\ref{fig:F-flucs}(c) for low temperature $k_BT\ll E_{g,2}^\mathrm{F}$, where the excitations freeze out. This is also revealed by the $T=0$ result in Fig.~\ref{fig:F-flucs}(d) which is independent of cell size.

The TL behaviour is observed as fluctuations increasing linearly with $T$ at large values of $T$ in Fig.~\ref{fig:F-flucs}(c), and by the fluctuations plateauing as $R$ increases in Fig.~\ref{fig:F-flucs}(d). 
The TL behaviour, as predicted in Table~\ref{tab:cellFlucLimits}, requires that both $T\gg E_{g,2}^\mathrm{F}/k_B\approx 0.56$ nK and $R\gg l_{\mathrm{F}}\equiv\hbar/\sqrt{ME_{g,2}^\mathrm{F}}\approx 6.1\,\mu$m, where $l_{\mathrm{F}}$ is the length scale associated with the F phase energy gap. 
We note that for $c_1>0$, the energy gap is always greater than zero, but for $c_1<0$ the energy gap goes to zero as the system approaches a transition point to the BA phase (see Fig.~3(c) of Ref~\cite{Kawaguchi2012R}), and the transverse spin susceptibility $\chi_{x,y}=1/E_{g,2}^\mathrm{F}$  and $l_{\mathrm{F}}$ diverge.\footnote{Similar behavior occurs at the F to AF phase boundary for $c_1>0$ (although the other magnon branch softens), and in the P to BA phase boundary for $c_1<0$. }

\subsection{P phase}

\begin{figure}[t]
\includegraphics[width=\linewidth]{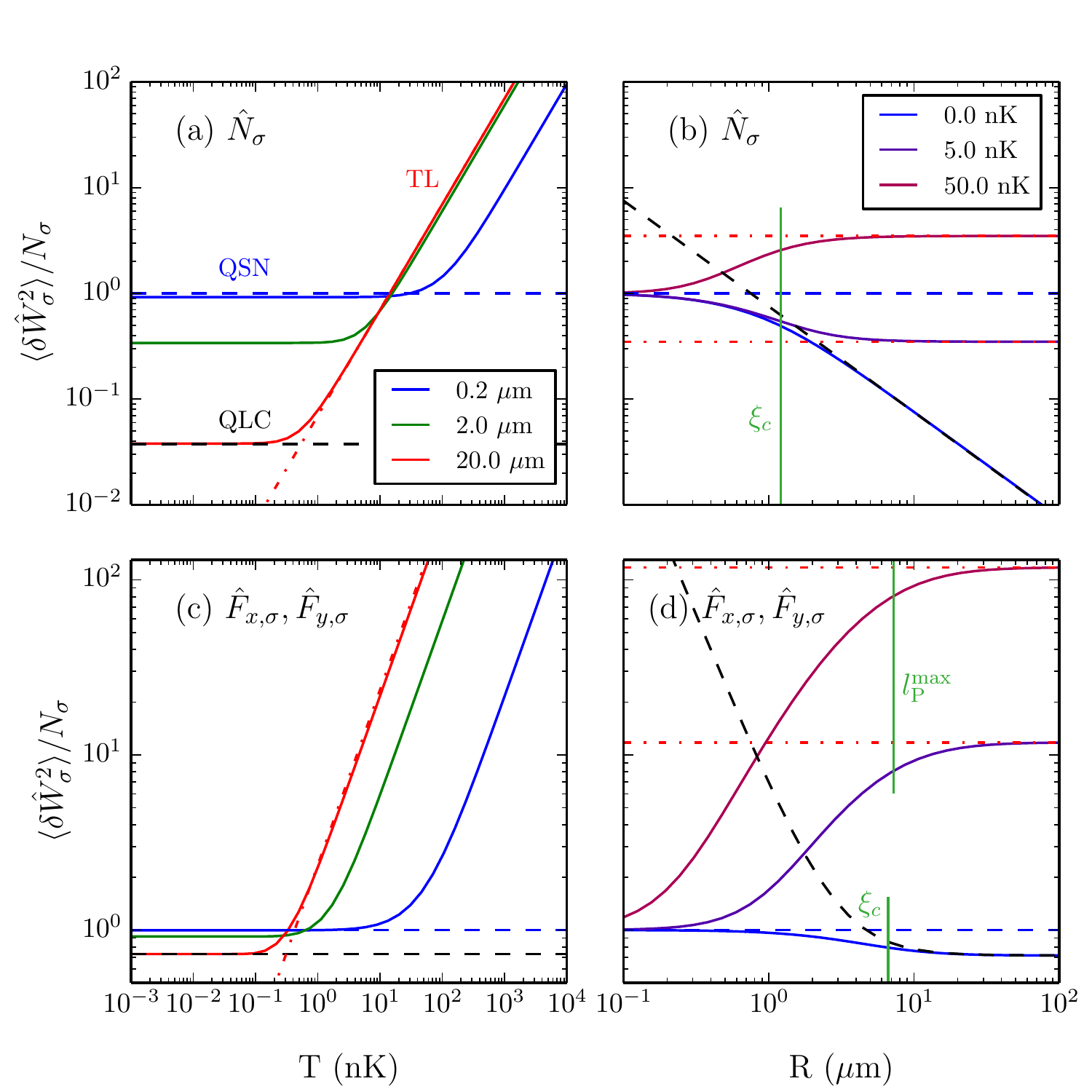}
\caption{
(Color online) Cell fluctuations for the P phase.
In (a) and (c) we plot temperature dependent results for
cell sizes of $R\in\{0.2, 2, 20\} \times \mu$m, as labelled in (a).
In (b) and (d) we plot fluctuations as a function of cell size $R$ for temperatures $T \in \{0, 5, 50\}$ nK, as labelled in (b).
Also plotted are the QSN limit (dashed blue line), QLC limit (dashed black line) and the TL (dot-dashed red line).
Parameters are for $^{23}$Na as in Table~\ref{tab:cellFlucParams}, with $f_z=0$, $p=1.5 c_1 n$ and $q=2.1 c_1n$.
For $\hat{N}_\sigma$, $\xic = 1.2$~$\mu$m and for $\hat{F}_{x,\sigma}$ and $\hat{F}_{y,\sigma}$, $\xic = 6.6$~$\mu$m
and $l_\mathrm{P}^\mathrm{max}$ are shown 
[vertical green lines in (b,d)].
}
\label{fig:P-flucs}
\end{figure}

Numerical results for the P phase are presented, along with analytic limits, in Fig.~\ref{fig:P-flucs} as functions of temperature $T$
and cell size $R$.

We note that, to our level of approximation [i.e.~Eq.~(\ref{e:dwBog})]  we have   $\Delta \hat{F}_{z,\sigma}^2=0$ in the P phase, since the condensate order parameter is $\boldsymbol{\xi}=[0,1,0]^T$. 
The $S_n$ structure factor for $\hat{N}_{\sigma}$  fluctuations arises from the phonon excitation branch. The general behavior of these fluctuations in Figs.~\ref{fig:P-flucs}(a) and (b) is similar to the analysis of  $\hat{N}_\sigma$ and $\hat{F}_{z,\sigma}$ fluctuations in the F phase, except that the value of $\xic$ is larger by $1\%$.

As in the F phase, the axial symmetry of the polar ground state means that the transverse magnetic fluctuations are isotropic  (i.e.~$\Delta F^2_{x,\sigma}  = \Delta F^2_{y,\sigma}$). 
The relevant structure factors $S_x(k)=S_y(k)$ are gapped and quadratic for low $k$. Both magnon excitation branches contribute to them, and they have a non-zero correlation length $\xic$.
This causes the QLC limit to exhibit non-extensive scaling [i.e.~dependence on $R$ for low temperature fluctuations in Fig.~\ref{fig:P-flucs}(c), and the QLC limit in Fig.~\ref{fig:P-flucs}(d)].   

At high temperatures we see similar behaviour to that observed for the transverse fluctuations in the F phase.
The two magnon excitation branches have energy gaps of  $E_{\mathrm{g},1}^{\mathrm{P}}=\sqrt{q(q+2c_1n)}- p$ and $E_{\mathrm{g},2}^{\mathrm{P}}=\sqrt{q(q+2c_1n)}+ p$, and obtaining TL fluctuations requires  $T\gg\max\{E_{\mathrm{g},1}^\mathrm{P},E_{\mathrm{g},2}^\mathrm{P}\}/k_B\!\approx 1.2$ nK and $R\gg l_{\mathrm{P}}^{\max}\equiv\hbar/\sqrt{M\min\{E_{\mathrm{g},1}^\mathrm{P},E_{\mathrm{g},2}^\mathrm{P}\}}\approx 7.2\,\mu$m, where $l_{\mathrm{P}}^{\max}$ is the longest length scale associated with the P phase energy gaps. 

\subsection{AF phase}\label{Sec:AF}

\begin{figure}[t]
 \includegraphics[width=\linewidth]{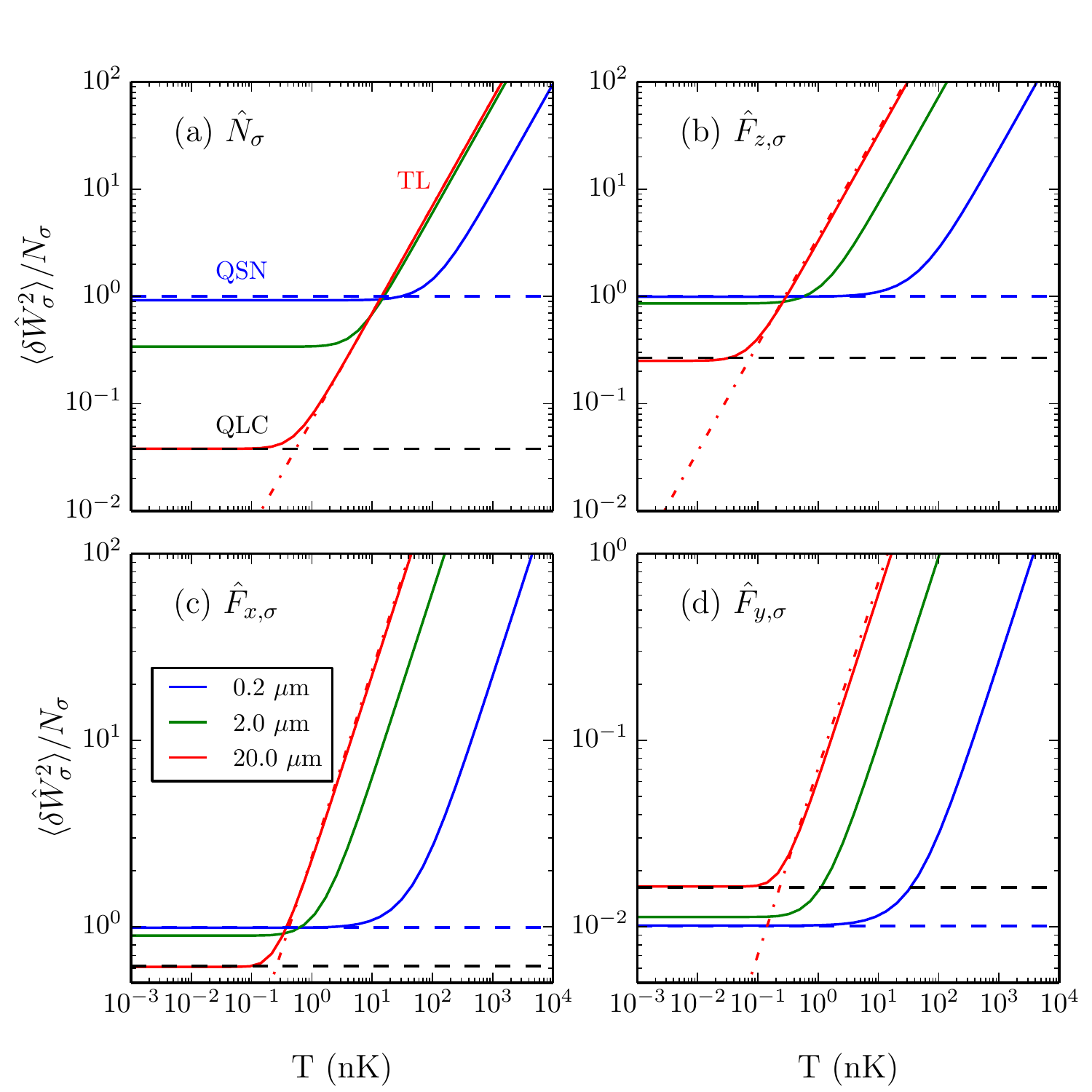}
 \caption{
(Color online) Cell fluctuations for the AF phase as a function of temperature, 
 for cell sizes $R \in \{0.2, 2, 20\} \mu$m as labelled in (c).
 Also plotted are the QSN limit (dashed blue line), QLC limit (dashed red line) and TL (dot-dashed red line).
 Parameters are for $^{23}$Na as in Table~\ref{tab:cellFlucParams}, with $f_z=0.2n$ and $q=-c_1n$.
 }
 \label{fig:AF-gaussian-flucs-vs-temp}
 \end{figure}

\begin{figure}[t]
 \includegraphics[width=\linewidth]{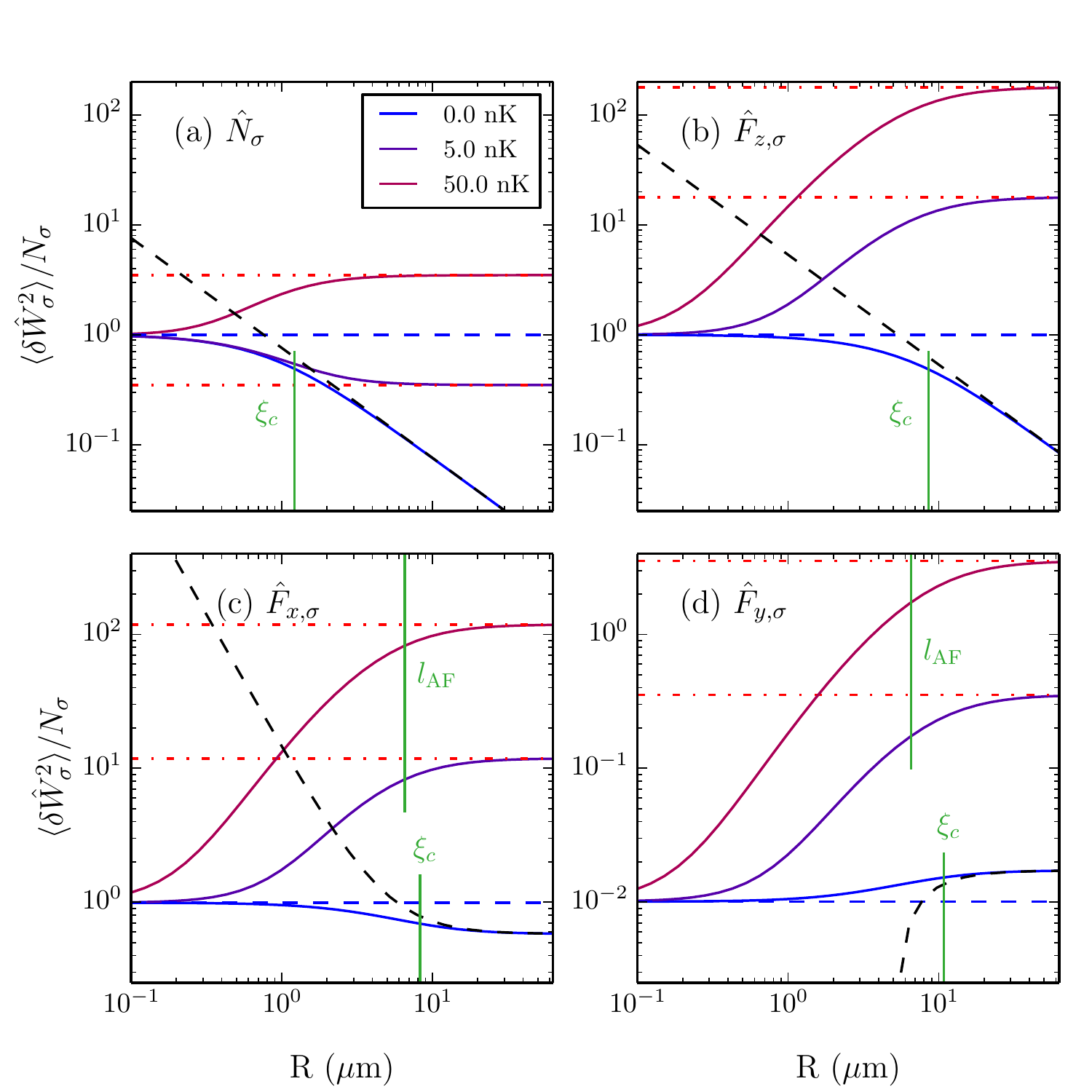}
 \caption{
(Color online)  Cell fluctuations for the AF phase as a function of cell size $R$, 
 for temperatures $T \in \{0, 5, 50\}$ nK as labelled in (a).
 Parameters, limits and line styles are as in Fig.~\ref{fig:AF-gaussian-flucs-vs-temp}.
For $\hat{N}_\sigma$, $\xic=1.2$~$\mu$m;
for $\hat{F}_{z,\sigma}$, $\xic=8.6$~$\mu$m;
for $\hat{F}_{x,\sigma}$, $\xic=8.3$~$\mu$m;
for $\hat{F}_{y,\sigma}$, $\xic=11$~$\mu$m;
are shown along with $l_\mathrm{AF}$ for $\hat{F}_{x,\sigma}$ and $\hat{F}_{y,\sigma}$ [vertical green lines].
 }
 \label{fig:AF-gaussian-flucs-vs-R-log}
 \end{figure}

Numerical results for the AF phase are presented, along with analytic limits,
in Fig.~\ref{fig:AF-gaussian-flucs-vs-temp} as a function of temperature $T$,
and in Fig.~\ref{fig:AF-gaussian-flucs-vs-R-log}
as a function of cell size $R$.

The behavior of $\hat{N}_{\sigma}$ and $\hat{F}_{z,\sigma}$ fluctuations are shown in Figs.~\ref{fig:AF-gaussian-flucs-vs-temp}(a),(b) and \ref{fig:AF-gaussian-flucs-vs-R-log}(a),(b). These results are similar to number fluctuations  of a scalar condensate, but with different correlation lengths (see Table~\ref{tab:cellFlucLimits}).  Notably, because the correlation length for $z$-spin fluctuations is larger, in Fig.~\ref{fig:AF-gaussian-flucs-vs-temp}(b) only the largest ($R=20\mu$m) cell is in good agreement with the TL prediction for fluctuations at large $T$. In the AF phase the axial symmetry is broken by the condensate order parameter having unequal nematic moments in the spin-$xy$ plane (e.g.~see order parameter in Fig.~\ref{Tab:phases}). Arising from this broken symmetry is a second Nambu-Goldstone mode: in addition to the phonon mode, a magnon mode develops that is gapless and has a linear dependence on $k$ as $k \to 0$. Both Nambu-Goldstone modes contribute to $\hat{N}_{\sigma}$ and $\hat{F}_{z,\sigma}$ fluctuations, however the phonon mode dominates  $\hat{N}_{\sigma}$, whereas the magnon mode dominates $\hat{F}_{z,\sigma}$ for $|f_z| \ll n$.

In Figs.~\ref{fig:AF-gaussian-flucs-vs-temp}(c),(d) and \ref{fig:AF-gaussian-flucs-vs-R-log}(c),(d) we consider the transverse spin fluctuations. The broken symmetry of the AF phase is revealed by the asymmetry in  $\hat{F}_{x,\sigma}$ and $\hat{F}_{y,\sigma}$ fluctuations. We note that by choosing the condensate order parameter to be real we bias the condensate nematic tensor to bulge out in the spin-$x$ direction relative to the  spin-$y$ direction (e.g.~see Sec.~3.3.2 of  \cite{Kawaguchi2012R}),  
(i.e.~$S_x^\infty>S_y^\infty$), which is reflected in the QSN limit for fluctuations [c.f.~Eqs.~(\ref{e:Sinfty}), (\ref{eq:quantumSmallCellLimit})], as can be seen by comparing the fluctuations at low $T$ for the  $R=0.2\mu$m cell in Figs.~\ref{fig:AF-gaussian-flucs-vs-temp}(c) and (d).

The low $k$ properties of the transverse spin structure factors also differ: $S_x$ has a dip [i.e.~$S_x(0)<S_x^\infty$, as illustrated in Fig.~\ref{fig:ModelStrucs}(b)], whereas $S_y$ has a peak. This difference changes the sign of the second term of the QLC limit (\ref{eq:QLgapped}), which governs the non extensive scaling of the fluctuations. This is particularly noticeable in the comparison of the low temperature plateaus in fluctuations in  Figs.~\ref{fig:AF-gaussian-flucs-vs-temp}(c) and (d): For $\hat{F}_{x,\sigma}$ quantum fluctuations get smaller (relative to $N_\sigma$) as cell size increases, whereas $\hat{F}_{y,\sigma}$ quantum fluctuations get larger. 

The TL for $\hat{F}_{x,\sigma}$ and $\hat{F}_{y,\sigma}$ fluctuations requires that $T\gg E_{g,2}^{\mathrm{AF}}/k_B\approx 0.49$ nK and $R\gg l_{\mathrm{AF}}\approx 6.6\,\mu$m, with these quantities defined in Table~\ref{tab:cellFlucLimits}. 

 We emphasise that in situations  where the axial symmetry is broken spontaneously, the behaviour we attribute to $x$ and $y$ spin fluctuations here will rotate to an arbitrary angle in the transverse plane.
  
\subsection{BA phase} \label{sec:BAcellFlucs}
\begin{figure}[t]
\includegraphics[width=\linewidth]{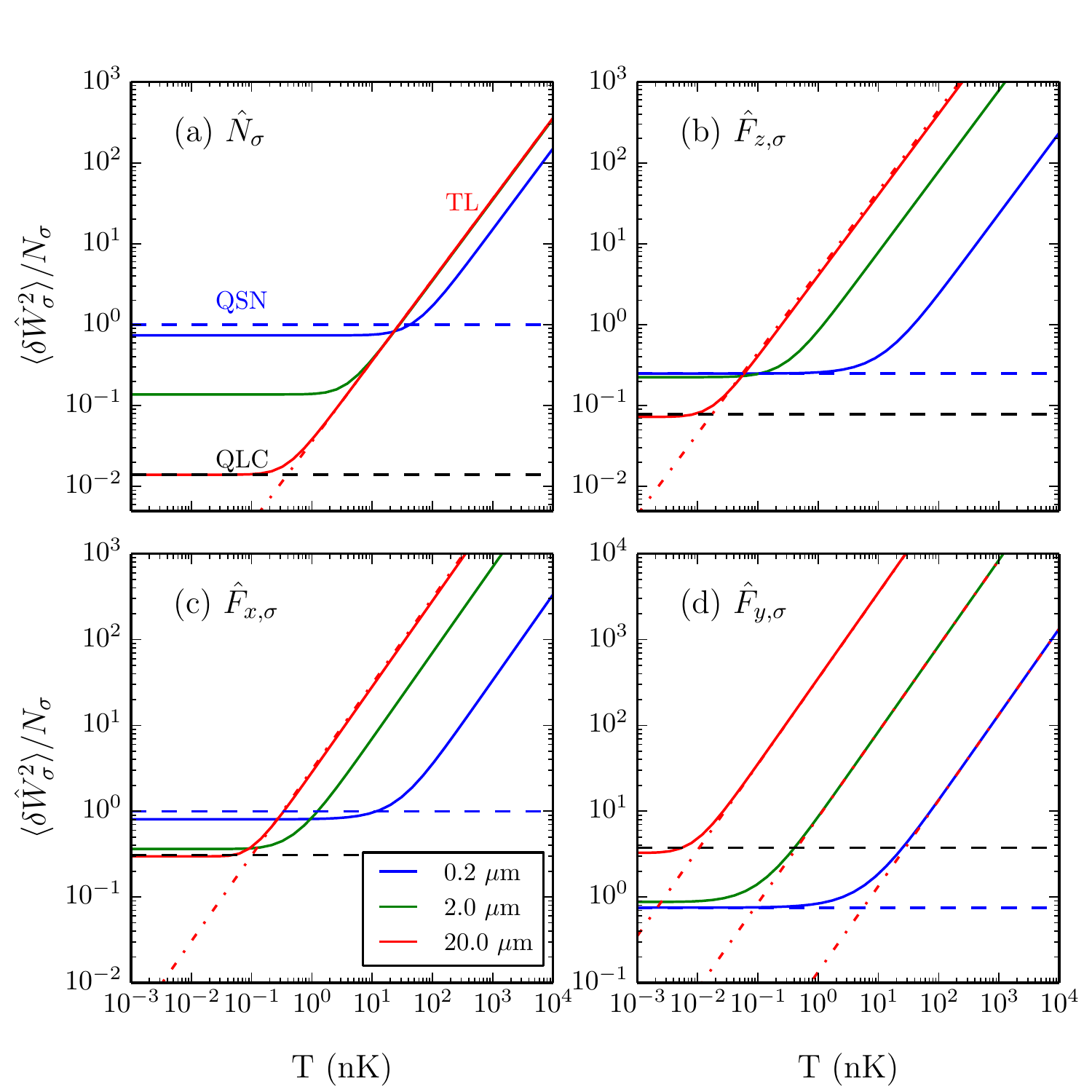}
\caption{
(Color online) Cell fluctuations for the BA phase as a function of temperature 
for cell sizes $R \in \{0.2, 2, 20\} \mu$m as labelled in (c).
Also plotted are the QSN limit (dashed blue line), QLC limit (dashed red line) and TL (dot-dashed red line).
Parameters are for $^{87}$Rb as in Table~\ref{tab:cellFlucParams}, with $f_z=0$, $p=0$ and $q=|c_1|n$.
To calculate $\hat{F}_{y,\sigma}$ fluctuations in (d) we use  $L_\text{sys} = 70 \, \mu$m (see text).
}
\label{fig:BA-gaussian-flucs-vs-temp}
\end{figure}

\begin{figure}[t]
\includegraphics[width=\linewidth]{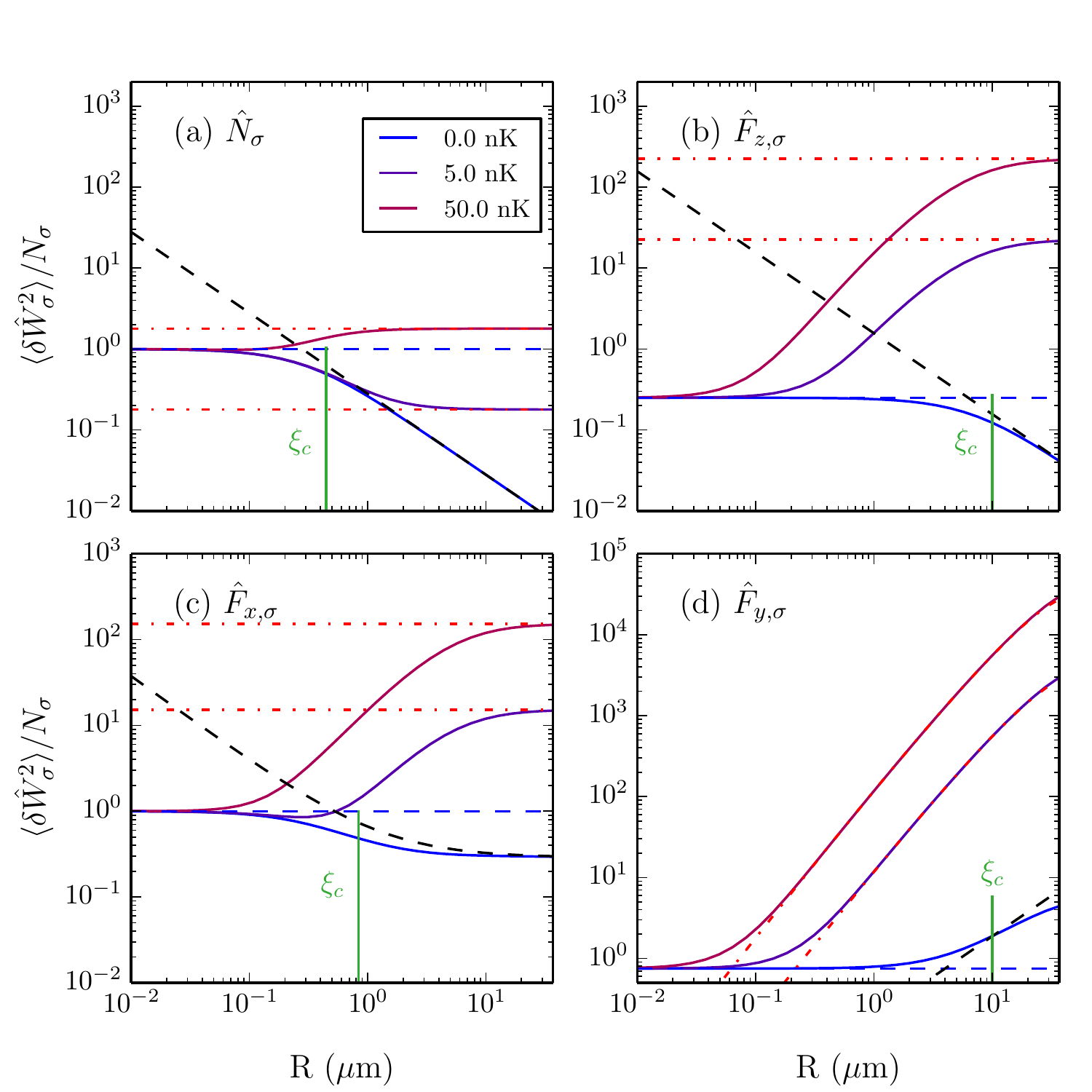}
\caption{
(Color online) Cell fluctuations for the BA phase as a function of cell size $R$,
for temperatures $T \in \{0, 5, 50\}$ nK as labelled in (a).
Parameters, limits and line styles are as in Fig.~\ref{fig:BA-gaussian-flucs-vs-temp}.
For $\hat{N}_\sigma$, $\xic = 0.45$~$\mu$m;
for $\hat{F}_{z,\sigma}$, $\xic = 10$~$\mu$m;
for $\hat{F}_{x,\sigma}$, $\xic = 0.84$~$\mu$m;
and for $\hat{F}_{y,\sigma}$, $\xic = 10$~$\mu$m [vertical green lines].
}
\label{fig:BA-gaussian-flucs-vs-R-log}
\end{figure}

Numerical results for the BA phase are presented,
along with analytic limits, in
Fig.~\ref{fig:BA-gaussian-flucs-vs-temp} as a function of temperature $T$,
and in Fig.~\ref{fig:BA-gaussian-flucs-vs-R-log}
as a function of cell size $R$. Note that this phase occurs for a negative spin dependent interaction ($c_1<0$) and we use parameters for  $^{87}$Rb to illustrate our results in this subsection.
We also restrict our attention to the BA phase at $p=0$, in which case the condensate spin is transverse and breaks the axial symmetry of the Hamiltonian.\footnote{For $p\ne0$ the condensate has an axial  spin component, but there  are no general analytic results for the condensate and excitations  (see~\cite{Murata2007a}).} This regime and the spontaneous broken symmetry have been explored in beautiful experiments by Sadler \textit{et al.}~\cite{Sadler2006a} where the condensate was quenched from the P phase to BA phase.  Due to the broken symmetry a Nambu-Goldstone magnon mode arises, which contributes to $\hat{F}_{z,\sigma}$  and $\hat{F}_{y,\sigma}$  fluctuations, notably causing $S_{y}$ to diverge (we discuss this further below).   

The $S_n$ and $S_z$ structure factors for the BA phase are both ungapped and linear at low $k$. The $\hat{N}_{\sigma}$ fluctuations arise from the phonon mode,\footnote{There is an avoided crossing with the gapped magnon mode, however for typical parameters of $^{87}$Rb  condensates this feature is negligible (see \cite{Symes2014a}).} whereas the $\hat{F}_{z,\sigma}$ fluctuations are due to the magnon Nambu-Goldstone mode. In both cases we observe that the fluctuations grow slower than $N_\sigma $ at $T=0$ [see Figs.~\ref{fig:BA-gaussian-flucs-vs-R-log}(a) and (b)]. Also, because the correlation length for the $\hat{F}_{z,\sigma}$  fluctuations is much larger than that for $\hat{N}_{\sigma}$, the cross over to the TL fluctuations for $\hat{F}_{z,\sigma}$ requires appreciably larger cells  [e.g.~see Figs.~\ref{fig:BA-gaussian-flucs-vs-temp}(a) and (b)].
 
The $T=0$ structure factor for  $\hat{F}_{x,\sigma}$ fluctuations is gapped, with a linear approach to its $k=0$ value.
Results for the fluctuations are shown in 
Figs.~\ref{fig:BA-gaussian-flucs-vs-temp}(c) and \ref{fig:BA-gaussian-flucs-vs-R-log}(c), and we see similar behaviour to transverse spin fluctuations in the P phase. 

For our choice of $^{87}$Rb parameters  the density healing length is appreciably smaller than for the sodium case considered previously (see Table \ref{tab:cellFlucParams}). For this reason the $0.2\:\mu$m cell  is not sufficiently small for the   QSN limit to hold [see Figs.~\ref{fig:BA-gaussian-flucs-vs-temp}(a) and (c)].

Finally, we consider  $\hat{F}_{y,\sigma}$ fluctuations.
Due to our choice of a real order parameter, the spontaneously acquired spin order is along $x$.
Thus fluctuations in the $y$-component of spin, which act to restore the axial symmetry, are dominated by the magnon Nambu-Goldstone mode.
We note from Ref.~\cite{Symes2014a} that the $T=0$ structure factor is divergent, and of the form given in Eq.~\eqref{eq:Sdiverge} with $j=-1$. Nevertheless in quasi-2D we
obtain a convergent QLC limit given by Eq.~\eqref{eq:QLdivergingSw}. This behavior is verified by the $T=0$ result Fig.~\ref{fig:BA-gaussian-flucs-vs-R-log}(d).\footnote{The disagreement at large $R$ is due to the introduction of $L_{\mathrm{sys}}$ in the numerical calculations [see discussion below Eq.~(\ref{eq:TinfSyBA_2})].}

In contrast, the $T>0$ structure factor diverges like  
\begin{align}
S_y(k \to 0) = S_y^\infty \left( \frac{ k l_T}{2} \right)^{-2} . \label{eq:TinfSyBA_2}
\end{align}
As might be expected for a spontaneous magnetization, the associated spin susceptibility $\chi_y$, and hence TL of the $\hat{F}_{y,\sigma}$ fluctuations, is divergent.
According to the Mermin-Wagner-Hohenberg (MWH) theorem, continuous symmetries cannot be spontaneously broken at finite temperature in a 2D system with short-range interactions  \cite{Wagner1966,Hohenberg1967}. In addition to forbidding the development of a long-range ordering of spin, this theorem also forbids the existence of a true condensate in 2D.

In experiments with finite trapped samples, a \textit{true condensate} can exist in a quasi-2D trap if the temperature is sufficiently cold for the phase coherence length to exceed the size of the system, $L_\mathrm{sys}$ \cite{Petrov2000}. Clearly the thermodynamic limit does not exist since the condensation temperature goes to zero as $L_\mathrm{sys}\to \infty$ (in accordance with the MWH theorem). However, this suggests that in any finite system where the theory is applicable, that the excitations will be physically cutoff by some length scale $L_\mathrm{sys}$. This length scale has not appeared explicitly thus far, but has been implicit in our assumption of a true condensate existing.\footnote{For the scalar case at higher temperatures a quasi-condensate phase exists with suppressed density fluctuations, but without off-diagonal-long-range order. However, the density fluctuations  of this phase are identical to  Bogoliubov predictions made assuming a true condensate (e.g.~see \cite{Mora2003a}).} 

The TL for $\hat{F}_{y,\sigma}$ fluctuations can be obtained by evaluating Eq.~(\ref{eq:wCellFlucsIntegral}) using Eq.~\eqref{eq:TinfSyBA_2}, adjusted to account for the long wavelength cutoff in quasi-2D: 
\begin{equation}
\Delta \W^{2}  = n\int_{1/L_{\mathrm{sys}}}^\infty  \frac{ k\,dk}{2\pi}\Sw(\mathbf{k})\tilde{\tau}_\sigma(\mathbf{k}).\label{eq:wCellFlucsIntegral2}
\end{equation} 
We find that the TL is logarithmically sensitive to $L_{\mathrm{sys}}$, and is given by
 \begin{align}
   \frac{\Delta  {F}_{y,\sigma}^2}{N_\sigma}
   &\approx S_y^\infty \left[ \ln\left(\frac{\sqrt{2}L_{\mathrm{sys}}}{R} \right) - \frac{\gamma_e}{2}\right] \left( \frac{2R}{l_T} \right)^2 , \label{eq:BAfyTLg}
\end{align}
where $\gamma_e \approx 0.5772$ is Euler's constant. This result is valid in the limit $l_T \ll R \ll L_\text{sys}$.
The QSN limit is insensitive to the long wavelength cutoff and is still given by our earlier result Eq.~(\ref{eq:quantumSmallCellLimit}).

Numerical results for $\hat{F}_{y,\sigma}$ fluctuations,
along with the analytic limits just derived, are shown in Fig.~\ref{fig:BA-gaussian-flucs-vs-temp}(d) as a function of temperature $T$,
and in Fig.~\ref{fig:BA-gaussian-flucs-vs-R-log}(d) as a function of cell size $R$, using a cutoff of the Thomas-Fermi radius of the system, i.e.~$L_{\mathrm{sys}} = 70\,\mu$m.
In the high $T$ regime, $l_\mathrm{T} \ll R$ and the fluctuations increase linearly with $T$.
Noting that $N_\sigma\propto R^2$ in quasi-2D,
Eq.~\eqref{eq:BAfyTLg} reveals that 
$ \Delta  {F}_{y,\sigma}^2 \propto N_\sigma^2$.\footnote{In the regime where the logarithmic term can be taken constant.} 
This is the remnant of the diverging susceptibility in the finite system, and is clearly seen in Fig.~\ref{fig:BA-gaussian-flucs-vs-R-log}(d) as the non-extensive scaling of the TL results, i.e.~fluctuations increasing rapidly with cell size. 
Similarly, the QLC result given by Eq.~\eqref{eq:QLdivergingSw} scales as $N_\sigma^{3/2}$, as can be seen in Fig.~\ref{fig:BA-gaussian-flucs-vs-R-log}(d).

\section{Conclusions}\label{Sec:Conclusion}

In this paper we have presented a theory for the fluctuations of atom number and components of spin in a spinor condensate. Our theory is based around the idea of Gaussian measurement cells which approximate finite resolution measurements that could be made in experiments, e.g., using dispersive \textit{in situ} imaging.  

Our results for the spin-1 system demonstrate how aspects of the magnetic phases are revealed by fluctuations in the various quantities considered. Importantly, the spin fluctuations are typically dominated by the magnon excitations of the system, and are sensitive to whether these are gapped or gapless. Notably, gapless Nambu-Goldstone magnons arise from spontaneously broken axial spin symmetry in the AF and BA phases. For the BA phase we showed that this leads to a divergence in a transverse spin component, which could be measured as a non-extensive scaling of the fluctuations. For the AF phase the equivalent divergence occurs in a component of the nematic density fluctuations. Our formalism naturally includes nematic fluctuations, although we did not present results for these in this paper. However, we note dispersive imaging techniques are also capable of measuring nematic density.
 
Another important feature of our results is the diverse range of length and temperature scales governing the relevant fluctuations. This occurs because the typical scales of density and spin interaction energies differ by  roughly two orders of magnitude in rubidium and sodium spinor condensates. Thus, while the density correlation lengths  are typically smaller than a micron and not resolvable with optical probing, the spin correlation lengths are much larger, and easily resolved. This will make it feasible to explore the role of cell size in fluctuation measurements.  

There are several exciting possibilities for future development of the ideas presented here. First, the extension to higher spin systems is immediate. Second, it would be interesting to include the role of dipole-dipole interactions. These interactions are quite weak for the case of $^{87}$Rb, but can have an appreciable role on the low energy spectrum, which would   affect   large-cell fluctuation measurements (also see \cite{Marti2014}). Finally, beyond the Bogoliubov approach we presented here there are many avenues for extending the theory including  the role of quantum and thermal back-action on the condensate  \cite{Phuc2011a,Kawaguchi2012a} and detailed behaviour of the system near phase boundaries (e.g.~see \cite{Huang2002a,Mukerjee2006a,Pietila2010a,Natu2011a,James2011a,Phuc2013a}).

\section*{Acknowledgments}
We acknowledge support by the Marsden Fund of New Zealand (contract UOO1220). 

\appendix
 
\bibliographystyle{apsrev4-1}
%\bibliography{spinor}


\end{document}